*Chapter 3*

# HIGH PERFORMANCE RECONFIGURABLE COMPUTING SYSTEMS

***Issam W. Damaj*** *

Dept. of Electrical and Computer Engineering Dhofar University

P.O. Box 2509, 211 Salalah, Oman

## Abstract

The rapid progress and advancement in electronic chips technology provides a variety of new implementation options for system engineers. The choice varies between the flexible programs running on a general purpose processor *(GPP)* and the fixed hardware implementation using an application specific integrated circuit (*ASIC*). Many other implementation options present, for instance, a system with a *RISC* processor and a *DSP* core. Other options include graphics processors and microcontrollers. Specialist processors certainly improve performance over general-purpose ones, but this comes as a quid pro quo for flexibility. Combining the flexibility of *GPPs* and the high performance of *ASICs* leads to the introduction of reconfigurable computing (*RC*) as a new implementation option with a balance between versatility and speed.

Field Programmable Gate Arrays (*FPGAs*), nowadays are important components of *RC*-systems, have shown a dramatic increase in their density over the last few years. For example, companies like *Xilinx* [1] and *Altera* [2] have enabled the production of *FPGAs* with several millions of gates, such as, the *Virtex-2 Pro* and the *Stratix-2 FPGAs*. Considerable research efforts have been made to develop a variety of *RC*-systems. Research prototypes with fine-grain granularity include *Splash* [3], *DECPeRLe-1* [4], *DPGA* [5] and *Garp* [6]. Examples of systems with coarse-grain granularity are *RaPiD* [7], *MorphoSys* [8], and *RAW* [9]. Many other systems were also developed, for instance, *rDPA* [10], *MATRIX* [11], *REMARC* [12], *DISC* [13], *Spyder* [14] and *PRISM* [15].

The focus of this chapter is introducing reconfigurable computers as modern supercomputing architectures. The chapter also investigates the main reasons behind the current advancement in the development of *RC*-systems. Furthermore, a technical survey of various *RC*-systems is included laying common grounds for comparisons. In addition, this chapter mainly presents case studies implemented under the *MorphoSys RC*-system. The selected case studies belong to different areas of application, such as, computer graphics and information coding. Parallel versions of the studied algorithms are developed to match the topologies supported by the *MorphoSys*. Performance evaluation and results analyses are included for implementations with different characteristics.

---

* E-mail address: i_damaj@du.edu.om;

# 1. Introduction

The rapid progress and advancement in electronic chips technology provides a variety of new implementation options for system engineers. The choice varies between the flexible programs running on a general purpose processor *(GPP)* and the fixed hardware implementation using an application specific integrated circuit (ASIC). Many other implementation options present, for instance, a system with a *RISC* processor and a *DSP* core. Other options include graphics processors and microcontrollers. Specialist processors certainly improve performance over general-purpose ones, but this comes as a quid pro quo for flexibility. Combining the flexibility of *GPPs* and the high performance of *ASICs* leads to the introduction of reconfigurable computing (*RC*) as a new implementation option with a balance between versatility and speed.

*GPPs* are programmed entirely through software. *GPPs* have wide applicability; nevertheless they may not match the computational needs of many applications. *ASICs* are custom designed for applications. The architecture of an ASIC exploits intrinsic characteristics of an applications algorithm that lead to a high performance. However, the direct architecture algorithm mapping restricts the range of applicability of ASIC-based systems. *ASICs* provide precise function needed for a specific task. The designer, by synchronizing each ASIC to execute a job, can produce chips that are fast, cheap and consume less power than programmable or general-purpose processors.

Along with trying to find the right balance between versatility and speed, designers have to face the cost constraint too. In addition to the cost of manufacturing several *ASICs*, another cost is the cost of design. Since a well-designed ASIC can solve a certain problem but not a slightly modified problem, efforts to design the new ASIC cannot make use of all the effort spent on the old ASIC since it is too highly customized to be reused. Thus, the effort expended on the design of an ASIC is almost lost when designing other ASIC, even those which performance of a closely related task.

With reconfigurable computing (*RC*), a new computation paradigm has emerged over the last decade, which intents to fill the gap between conventional microprocessors and application-specific integrated circuits (*ASICs*) [16]. All reconfigurable systems share the same basic idea: to benefit from programmable logic, which allows to dynamically adapting the system's functionality to the requirements of the running application. The most popular devices, which actually enabled reconfigurable computing, are Field-programmable Gate Arrays (*FPGAs*), which were introduced in the mid eighties. Many approaches of reconfigurable systems have been proposed in recent years; some of them are known as hybrids, which combine reconfigurable hardware with a processor core such as the *MorphoSys* reconfigurable system designed at the University of California Irvine (*UCI*).

The focus of this chapter is introducing reconfigurable computers as modern supercomputing architectures. The chapter also investigates the main reasons behind the current advancement in the development of *RC*-systems. Furthermore, a technical survey of various *RC*-systems is included laying common grounds for comparisons. In addition, this chapter mainly presents case studies implemented under the *MorphoSys RC*-system. The selected case studies belong to different areas of application, such as, computer graphics, information coding, and signal processing. Parallel versions of the studied algorithms are developed to match the topologies supported by the *MorphoSys*. Performance evaluation and results analyses are included for implementations with different characteristics.

# 2. Reconfigurable Computing Systems

## 2.1. History of Reconfigurable Computing

The first descriptions of computing automata capable of reconfiguration were put forward by John von Neumann in a series of lectures and unfinished manuscripts dating back to the late 40's and early 50's. After the death of von Neumann in 1957, his works on self-reproducing automata were collected and edited by Arthur Burks, who published them in 1966. Although von Neumann is generally regarded as the main developer of the conventional serial model of computing, it seems obvious that during the last years of his life, he was more interested in more complicated computing automata. The earliest electronic computers were built with error-prone vacuum tube technology. In 1959, Jack Kilby invented the monolithic integrated circuit at Texas Instruments, but it was not until the introduction of Intel's 4004 microprocessor that general-purpose computers began to be integrated on the same silicon chip.

In 1963, Gerald Estrin of University of California at Los Angeles proposed a variable structure computer system to achieve performance gains in a variety of computational asks [18]. The central idea was to combine both fixed and variable structure computer organizations, where the variable subsystem could be reorganized into a variety of problem-oriented special purpose configurations.

The first suggestion for a programmable logic device is due to Sven Wahlstrom, who in 1967 proposed the inclusion of additional gates to customize an array of integrated circuitry. However, the silicon "real estate" was an extremely scarce resource in those days. In 1967, Robert Minnick published a survey of microcellular research. He described both fixed cell-function arrays and variable cell-function arrays. In fixed cell-function arrays the switching function of each cell remained fixed, and only the interconnections between cells were programmable. In the case of variable cell-function arrays, the function produced by each cell could also be determined by parameter selection.

In the 70's, interest in and the corresponding financial support for non-serial forms of computation seems to have tapered off. This was most probably caused by the introduction of the first microprocessor - the famous 4004 by Intel - in 1971 and the ever-growing number and scope of applications enabled by the expanding market of microprocessors and microcontrollers.

In the eighties, there was a revived interest in both systolic and parallel architectures. This interest was inspired partly by the advances in semiconductor integration technology and the evolution of system design concepts —the seminal work on Very Large-Scale Integration (*VLSI*) design by Carver Mead and Lynn Conway was published in 1980 and partly by new and more demanding applications of supercomputing. An interesting design combining parallelism with reconfigurability was the Texas Reconfigurable Array Computer (*TRAC*) [19]. In the *TRA* project, reconfigurability meant the reprogramming of interconnections between individual computing elements.

In the late eighties and early nineties, the first platforms for reconfigurable computing were built. One of the first such platforms was designed at Digital Equipment Corporation's (*DEC*) Paris Research Laboratory (*PRL*) and was called Programmable Active Memory (*PAM*) [20]. The speedups achieved by the *PAM* project were very impressive and as similar results were reported by other research groups at approximately the same time, one could say that reconfigurable computing had passed its first test. This was also realized by two influential engineering societies, the Association for Computing Machinery (*ACM*) and the Institute of Electrical and Electronics Engineers (*IEEE*). These engineering societies began sponsoring two annual conference series about the applications of *FPGAs*, namely the *ACM/SIGDA* International Symposium on Field-Programmable Gate Arrays and the *IEEE* Symposium on *FPGA*-Based Custom Computing Machines. In Europe, the first International Workshop on Field-Programmable Logic and Applications was held in 1991. Reconfigurable computing would not be possible without the advances in electronics, because large *FPGA* circuits have enormous silicon overhead; for example, a programmable logic device with 50000 usable gates may have well over a million transistors. This demonstrates that the *FPGA* market has benefited tremendously from advances in semiconductor manufacturing technology.

*Xilinx*, which was founded in 1984, introduced the world's first *FPGA* in 1985. Being the first in the *FPGA* market, Xilinx continued to dominate it well into the nineties, but in recent years *Altera* Corporation has increased its market share substantially, due to the popularity of its new *SRAM* -based *FPGAs*, the *FLEX* 8000 and *FLEX* 10K family. The convergence of the theoretical path and the technological path in the nineties may mark the beginning of a promising era for reconfigurable computing. Many influential industry observers feel, that the promises of reconfigurable computing are not just superficial media hype, but that there are real advantages to be gained by applying reconfigurable computing.

## 2.2. Field Programmable Gate Arrays

*FPGAs* are a hybrid device between *PALs* and Mask-Programmable Gate Arrays (*MPGAs*). Like *PALs*, they are fully electrically programmable, and they can be customized nearly instantaneously. Like *MPGAs* they can implement very complex computations on a single chip, with millions of gates devices currently in production. These devices have opened up completely new avenues in high-performance computation, forming the basis of reconfigurable computing. Most current *FPGAs* are *SRAM* -programmable. This means that *SRAM* bits are connected to the configuration points in the *FPGA* and programming the *SRAM* bits configures the *FPGA*. Thus, these chips can be programmed and reprogrammed as easily as a standard static *RAM*. Here the programming bit will turn on a routing connection when it is configured with a true value, allowing a signal to flow from one wire to another, and will disconnect these resources when the bit is set to false. With a proper interconnection of these elements, which may include millions of routing choice points within a single device, a rich routing network can be created.

Xilinx, one of the major manufacturers of *FPGAs* defines them as high-density Application Specific Integrated Circuits (*ASICs*) combining the logic integration of custom Very Large Scale Integration (*VLSI*) with the time to market and cost advantages of standard products [1].

Another major producer *Lucent* also stresses the versatility of the *FPGA* in its publication and of the time to market benefits. In the beginning *FPGAs* were mostly looked at as being of importance for rapid prototyping of circuits which would later be hardwired and thus of being a design tool to aid profitability rather than being of use directly in applications. That is changing as chips become available with higher gate counts on tools and faster routing and placing tools becoming available [21].

Utilizing high-level programming languages allows a complete picture of structure and functionality of the whole circuit is built up via a text file that is then fed through the appropriate software to produce a layout of gates, known as a net list, and downloaded to the *FPGA*. This is where the advantages of a system based around *FPGAs* for rapid prototyping is seen.

## 2.3. Reconfigurable Systems Generalities

Reconfigurable hardware can be used to provide reconfigurable functional units within a host processor. This allows for a traditional programming environment with the addition of custom instructions that may change over time. The reconfigurable units execute as functional units controlled by a main microprocessor. Registers are used to hold the input and output operands. Thus, the basic architecture of comprises a software programmable core processor and a reconfigurable hardware component. The core processor executes sequential tasks of the application and controls data transfers between the programmable hardware and data memory. Generally, the reconfigurable hardware is dedicated to exploitation of parallelism available in the applications algorithm. This hardware typically consists of a collection of interconnected reconfigurable elements. Both the functionality of the elements and their interconnection is determined through a special configuration program called the context.

State of the art *RC*-systems are of different architectures. A general classification could be viewed in terms of 4 main categories: granularity, depth of programmability, reconfigurability, and interface coupling. We note at this point that more detailed taxonomies are presented in [22, 23]. These 4 main categories are defined as follows:

### 2.3.1. Granularity

System granularity is defined by the internal structure of the reconfigurable elements. Computation blocks within the reconfigurable hardware vary from system to system. Each unit of computation can be as simple as a 3-input look up table (*LUT*), or as complex as a 4-bit *ALU*. This difference in block size is commonly referred to as the granularity of the logic block. A 3-bit *LUT* is an example of a very fine-grained computational element, and a 4-bit *ALU* is an example of a quite coarse-grained unit. Each element operates at the bit level implementing a Boolean function or a finite-state machine. The finer grained blocks are useful for bit-level manipulations, while the coarse-grained blocks are better optimized for standard datapath applications. Several reconfigurable systems use a medium-sized granularity of logic block. A number of these architectures operate on two or more 4-bit wide data words. Examples of fine-grain reconfigurable systems are *Splash* and *DECPeRLe-1*, and *Matrix* is an example of a coarse-grained reconfigurable system.

### 2.3.2. Depth of Programmability

In terms of depth of programmability, a reconfigurable system may have a singlecontext or multiplecontexts. For single-context systems only one configuration program context may be resident in the system. In this case the systems functionality is limited to the context currently loaded. On the contrary, in multiple-context systems, several contexts can be resident in the system at once. This allows execution of different tasks simply by changing the operating context.

### 2.3.3. Reconfigurability

Reconfigurability pertains to the ability of the system to overlap execution with loading with new context. In statically reconfigurable systems, reconfiguration of the programmable hardware can occur only if the current execution is interrupted or when it finishes. On the other hand, in dynamically reconfigurable systems reconfiguration can be done concurrently with execution. The interface coupling of a reconfigurable system refers to the level of integration of the core processor and the reconfigurable hardware.

Frequently, the areas of a program that can be accelerated through the use of reconfigurable hardware are too numerous or complex to be loaded simultaneously onto the available hardware. For these cases, it is helpful to use dynamically *RC*-systems to swap different configurations in and out of the reconfigurable hardware as they are needed during program execution. Accordingly, the run-time reconfigurability is more likely to lead to an overall improvement in performance.

### 2.3.4. Interface Coupling

An *RC*-system is tightly coupled if the core processor and the programmable component reside in the same chip. The system is loosely coupled, if core processor and programmable logic are implemented as separate devices.

In loosely coupled systems, an attached reconfigurable processing unit behaves as if it is an additional processor in a multiprocessor system. The host processor's data cache is invisible to the attached reconfigurable processing unit. Thus, a higher delay exists in communication between the host processor and the reconfigurable hardware. Such as, when communicating configuration information, input data, and results. However, this type of reconfigurable hardware does allow for a great deal of computation independence, by shifting large chunks of a computation over to the reconfigurable hardware. The most loosely coupled form of reconfigurable hardware is that of an external standalone processing unit. This type of reconfigurable hardware communicates infrequently with a host processor. This model is similar to that of networked workstations, where processing may occur for very long periods of time without a great deal of communication.

Each of the addressed styles has distinct benefits and drawbacks. The tighter the integration of the reconfigurable hardware, the more frequently it can be used within an application or set of applications due to a lower communication overhead. The more loosely coupled styles allow for greater parallelism in program execution but suffer from higher communications overhead.

## 2.4. Reconfigurable Systems

There has been considerable research effort to develop a variety of *RC*-systems. Research prototypes with fine-grain granularity include *Splash* [3], *DECPeRLe-1* [4], *DPGA* [5] and *Garp*[6]. Array processors with coarse-grain granularity, such as *rDPA* [10], *MATRIX* [11], and *REMARC* [12] form another class of reconfigurable systems. Other systems with coarse-grain granularity include *MorphoSys* [25-27], *RaPiD* [7], and *RAW* [9]. Other reconfigurable systems with a core control processor and *FPGAs* as the reconfigurable part are *DISC* [13], *Spyder* [14], and *PRISM* [15]. Other systems include the *PipeRench* [24].

The *Splash* and *DECPeRLe-1* computers were among the first research efforts in reconfigurable computing. *Splas*h, a linear array of processing elements with limited routing resources, is useful mostly for linear systolic applications. *DECPeRLe-1* is organized as a two-dimensional array of 16 *FPGAs* with more extensive routing. Both systems are fine-grained, with remote interface, single configuration and static reconfigurability.

*rDPA*, The reconfigurable data-path architecture (*rDPA*) consists of a regular array of identical data-path units (*DPUs*). Each *DPU* consists of an *ALU*, a micro-programmable control and four registers. The *rDPA* array is dynamically reconfigurable and scalable. The *ALUs* are intended for parallel and pipelined implementation of complete expressions and statement sequences. The configuration is done through mapping of statements in high-level languages to *rDPA* using *DPSS* (Data Path Synthesis System).

*MATRIX* is an array of 8-bit basic units (*BUs*), *ALU*-multiplication unit and control logic interconnected through a hierarchy of three levels. *MATRIX* aims to unify resources for instruction storage and computation. The basic unit (*BU*) can serve either as a memory or a computation unit. The 8-bit *BUs* are organized in an array, and each *BU* has a 256-word memory, *ALU*-multiply unit and reduction control logic. The interconnection network has a hierarchy of three levels, and it can deliver up to 10 *GOPS* (Giga-operations/s) with 100 *BUs* when operating at 100 *MHz*.

*REMARC* consists of a reconfigurable coprocessor, which has a global control unit for 64 programmable blocks (nano-processors). Each 16-bit nano-processor has a 32 entry instruction *RAM*, a 16-bit *ALU*, 16 entry data *RAM*, instruction register, and several registers for program data, input data and output data. The interconnection is two-level (*2D* mesh and global buses across rows and columns). The global control unit (1024 instruction *RAM* with data and control registers) controls the execution of the nano-processors and transfers data between the main processor and nano-processors. This system performs well for multimedia applications, such as *MPEG* encoding and decoding (though it is not specified if it satisfies the real-time constraints).

*RaPiD* is a linear array (8 to 32 cells) of functional units, configured to form a linear computation pipeline. Each array cell has an integer multiplier, three *ALUs*, registers and local memory-segmented buses are used for efficient utilization of interconnection resources. It achieves performance close to its peak 1.6 *GOPS* for applications such as *FIR* filters or motion estimation.

The Reconfigurable Architecture Workstation *(RAW*) is a set of replicated tiles, where each tile contains a simple *RISC* processor, some bit-level reconfigurable logic and some memory for instructions and data. Each *RAW* tile has an associated programmable switch, which connects the tiles in a wide-channel point-to-point interconnect. When tested on benchmarks ranging from encryption, sorting, to *FFT* and matrix operations, it provided gains up to 100 times, as compared to a *Sun SparcStation* 20.

The Dynamically Programmable Gate Arrays (*DPGA*) is a fine-grained prototype system that use traditional 4-input lookup tables as the basic processing elements. *DPGA* supports rapid run-time reconfiguration.

*Garp* is a loosely coupled system with fine granularity. *Garp* has rows of blocks, which are like the *CLBs* of the *Xilinx* 4000 *FPGA* series. *Garp* architecture has more than 24 columns of blocks, whilst the number of rows is implementation dependant. The blocks operate on 2-bit data. There are vertical and horizontal block-to-block wires for data movement within the array. Separate memory buses move information (data as well as configuration) in and out of the array. Speedups ranging from 2 to 24 times are obtained for applications, such as encryption, image dithering and data sorting.

*MorphoSys*: Morphing System features a novel architecture for reconfigurable computing systems. The *MorphoSys* is primarily targeted to applications with inherent parallelism, high regularity, word-level granularity, and computations with intensive nature. Some examples of such applications are video compression, image processing, graphics acceleration, and security.

The first Processor Reconfiguration through Instruction-Set Metamorphosis (*PRISM-I*) consists of a board with four *Xilinx* 3090's plugged into a host system based around a *Motorola* 68010 system. A *C*-language-like compiler was created for *PRISM-I* to automatically translate subroutines to be mapped onto the reconfigurable hardware. Compiled programs run partly on the host processor and partly on the attached *FPGA*. The second prototype, *PRISM-2*, brought the host processor and *FPGAs* closer together, attaching an *AMD Am29050* directly to three *Xilinx* 4010's.

*Spyder* extended a custom processor with three *Xilinx* 4010 *FPGAs* acting as reconfigurable execution units. With *Spyder*, the programmer is responsible for dividing a program between the main processor and the reconfigurable units and programming each in a special subset of *C++* programming language.

*DISC* - the Dynamic Instruction Set Computer constructs the main processor and reconfigurable component together within the *FPGA* parts. The first *DISC* was made with two *National Semiconductor CLAy31s FPGAs*. A primitive main processor is implemented on a part of one *CLAy31*, with the majority of the same chip supplying the prototype reconfigurable component. The second *CLAy31* served only to control the loading of configurations on the first one. With *DISC-2*, the main processor is moved onto a separate third *CLAy31*.

*PipeRench* is a reconfigurable fabric - an interconnected network of configurable logic and storage elements. By virtualizing the hardware, PipeRench overcomes the disadvantages of using *FPGAs* as reconfigurable computing fabrics. Unlike *FPGAs*, *PipeRench* is designed to efficiently handle computations. Using a technique called pipeline reconfiguration; *PipeRench* improves compilation time, reconfiguration time, and forward compatibility.

On the industrial level, many *RC*-systems are currently being produced by different companies like *Xilinx* [1], *Celoxica* [28], *Elixent* [29], *Altera* [2], *Lucent* [21], *Actel* [30], *NallaTech* [31], *Chameleon Systems* [32], *MorphoTech* [33], and *Intel* [34].

## 2.5. Application of Reconfigurable Systems

Reconfigurable computers where brought in to many areas of application, such as, information coding, digital signal processing, digital image processing, space and solar applications, biomedical engineering, networking, and computers and communications security.

### 2.5.1. Information Coding

Reconfigurable hardware implementation of the *Viterbi* coding algorithm is presented in [35]. The investigation includes the implementation of a reduced-complexity adaptive *Viterbi* algorithm (*AVA*). Run-time dynamic reconfiguration is used in response to changing channel noise conditions to achieve improved decoder performance. Implementation parameters for the decoder have been determined through simulation. The decoder has been implemented on a *Xilinx* XC4036-based *PCI* board. An overall decode performance improvement of 7.5 times for *AVA* has been achieved versus algorithm implementation on a *Celeron*-processor based system. The uses of dynamic reconfiguration lead to a 20 percent performance improvement over a static implementation with no loss of decode accuracy.

### 2.5.2. Space and Solar Applications

For solar applications, the research done in [36] describes how the advent of *FPGAs* has allowed replacement of a huge number of *DSP* chips (30,000) by a smaller number (231) of custom chips with *Nobeyama* antenna array for monitoring solar flares.

In [37] N-body methods are used to simulate the evolution and interaction of galaxies using reconfigurable computing. These simulations are usually run on large-scale supercomputers or on very expensive full-custom reconfigurable hardware.

### 2.5.3. Digital Signal Processing

Considerable research for digital signal processing (*DSP*) has been done, such as; *RC* hardware implementations for elliptic *2R* filters, discrete Fourier transform (*DFT*), *FIR* filters, and adaptive digital filters.

The use of *FPGAs* to implement a fourth order band pass elliptic 2R filter is investigated in [38]. The key idea is to replace multipliers with Distributed Arithmetic (*DA*), in order to simplify the process involved. *DA* is where numbers are converted to two's complement digital format and then the multiplier can be implemented using adders and accumulators.

In [39], the investigation describes a new approach to computing the Discrete Fourier Transform (*DFT*) that significantly increases the power of *DSP* chips. The *DFT* is reduced to additions performed in parallel register arrays ideal for implementation in *FPGAs*. The algorithm is highly flexible and 10 to 100 times faster than standard approaches due to its elimination of add, multiply and accumulate operations.

*FIR* filters implementations where carried in different research groups [40 - 43] and industrial firms (e.g. Xilinx). Some of that work was done in the University of Kansas for synthesizing an efficient *FIR* filter reconfigurable architecture using *FPGAs*. The *FPGA* implementation suggests that 60-70 tap chips with sampling rates exceeding 100 MHz should be feasible [42]. Another automatic implementation of *FIR* filters on *FPGAs* is given in [43].

In [44] the application of reconfigurable computing to exploit the inherent serial features in some *DSP* algorithms is investigated with the aim of building cost-effective, real-time hardware. The addressed *DSP* algorithms are decomposed so that the most computationally-intensive and data-flow-oriented part is separated from the less intensive and more control-flow-oriented part. The different modules in the intensive part are expected to be serialized and implemented on a reconfigurable platform. The proposed architecture consists of an array of a minimum of two Field Programmable Gate Arrays (*FPGAs*). The *FPGAs* are grouped in two sets such that when one set executes the current batch of modules the second set could be configured to execute the next batch of modules. The control flow oriented part of the algorithm is implemented on a *DSP* processor.

Research is also carried for the creation of adaptive digital filters implemented in *FPGAs*, with the attraction that the coefficients and the filter structure can be changed through the reconfiguration of the *FPGA* [45].

### 2.5.4. Digital Image Processing

Accelerating desktop publishing (*DTP*) with Reconfigurable Computing Engines is discussed in [46]. This research deals with how a reconfigurable computing engine can be used to accelerate *DTP* functions. Also, how *PostScript* rendering can be accelerated using a commercially available *FPGA* co-processor cards. In the case of the *FPGA PostScript* project, *Xilinx* XC6200 *FPGA* is used to accelerate the computationally intensive areas of *PostScript* rendering. Efforts were done for the development of plug-ins for *Adobe Photoshop* which use the same board to accelerate image processing operations like color space conversion and image convolution.

In [47], the authors describe the use of the *SPLASH-2* custom computing platform for real-time median and morphological filtering images. *SPLASH-2* is an *FPGA*-based attached processor that can be reconfigured to perform a wide variety of tasks. Although not specifically designed for image processing, the architecture is well suited for the repetitive computations and high data transfer rates that characterize most low-level image processing problems. Median filtering is a particularly good benchmark, since nonlinear rank ordering must be performed for *2D* neighborhoods at every pixel location in an image. General-purpose workstations are inefficient at such tasks, whereas *SPLASH-2* is configured to perform this at a rate of 30 images per second. This research presents the hardware/software co-design process that have been used to implement this operation, which can be pipelined with other operations by using additional *SPLASH-2* processor boards.

A prototype end-to-end real time video coded is described in [45]. The presented system is implemented on a *CLAy31* (Configurable Logic Array) by National Semiconductors. One of the steps decomposes the image using a discrete wavelet transform (*DWT*), whose filters consume many resources of the *FPGA*. The rest of the chip is used by the associated addressing and control logic and frame grabber interface.

Another example of partial reconfiguration in video processing applications is presented in [48]. The hardware platform used in this case was the Field Configurable Multi-Chip Module (*FCM*) by National Semiconductor.

### 2.5.5. Biomedical Engineering

Reconfigurable computing is also used in the field of biomedical engineering. Some efforts were spent on a high-speed, voxel data processing computer architecture. A high-performance computer architecture aimed at computationally intensive applications of biomedical volume data processing is proposed in [49]. Algorithms used on volumetric data include visualization and digital radiograph reconstruction by ray-casting, oncological dose calculation, *3D FFT*, and *3D*-convolution processing. The generation of voxel address for algorithms that perform voxel data manipulation using a special address generator *FPGA*.

### 2.5.6. Networking

Some research in networking applications with reconfigurable computing is done for examining the possibility of introducing the *MorphoSys* reconfigurable system into *IP* routing. *IP* routing algorithms are mapped onto this system and its performance evaluated in order to measure the efficiency of this system relative to existing *IP* routing implementation, mainly, multi-homed computers and dedicated physical routers. The results were obtained by running the emulator program *mULATE* to implement and evaluate the performance of *MorphoSys*. Results have indicated that MorphoSys presented a much better performance than both multi-homed computers and dedicated physical routers, and can therefore be integrated as an active element in *IP* routing.

The research in [50] introduces a suite of tools called *NCHARGE* (Networked Configurable Hardware Administrator for Reconfiguration and Governing via End-systems). The system has been developed to simplify the co-design of hardware and software components that process packets within a network of *FPGAs*. A key feature of *NCHARGE* is that it provides a high-performance packet interface to hardware and standard Application Programming Interface (*API*) between software and reprogrammable hardware modules. Using this *API*, multiple software processes can communicate to one or more hardware modules using standard *TCP/IP* sockets. *NCHARGE* also provides a Web-Based user interface to simplify the configuration and control of an entire network switch that contains several software and hardware modules.

### 2.5.7. Security

In recent years, we have witnessed a rapid increase in the number of individuals and organizations using advanced data communications and computer networks for personal and professional activities. Among the variety of new uses of data communications, there are several applications which are highly sensitive to data security. Examples are commercial exchange on the Internet, computer networks, wireless communications, and military.

Reconfigurable devices such as *FPGAs* are a highly attractive option for hardware security, mainly for cryptographic algorithms. They provide the flexibility of a dynamic system as well as the ability to easily implement a wide range of algorithms [51]. Potential advantages of encryption algorithms implemented in *FPGAs* include the following:

Algorithm Agility: This expression refers to the change of the used cryptographic algorithm during operation. For example, during a session in one of the modern security protocols like *SSL*, an exchange could be done between *3DES*, *Blowfish*, *IDEA*, or any other algorithm. As they are reprogrammable, *FPGAs* seems to be a cheap alternative for traditional hardware with such a continuously evolvable area [52].

Algorithm Upload: It is obvious that *FPGAs* could be upgraded with a new cipher that did not exist (or was not standardized) at design time. Algorithm Modification: The modification of a standardized security algorithm is possible, for instance, by using S-boxes, permutations, or even as changing the mode of operation. Such modifications could be easily done with a reconfigurable hardware.

Architecture Efficiency: In certain cases, hardware architecture can perform better if it is designed for a specific set of parameters. For example, with fixed keys the main operation in the *IDEA* cipher [53] degenerates into a constant multiplication which is far more efficient than a general multiplication. Using *FPGAs* enables the switching to much more efficient implementation in certain specified cases.

Throughput: Although typically slower than *ASIC* implementations, *FPGA* implementations have the potential of running substantially faster than software implementations.

Reconfigurable hardware security has been the aim of many research investigations. Ploog et al studied in [54] about how modern smartcards can perform high security operations. The authors emulated the main smartcard algorithms in *FPGA* hardware and quantified the impact of the main *ASIC* design parameters on overall speed and silicon area.

Kim et al presented in [55] the use of *FPGAs* for a fully-pipelined, 56-bit *DES* encryption (decryption) and authentication at memory-bus bandwidths. Other implementation for *DES* was presented by Tom Kean and Ann Duncan from *Xilinx* in [56]. Prototype designs were realized and tested on the *XC6200DS PCI* Development System.

The *IDEA* was addressed in [53] by Davor and Mario presenting an *FPGA* core implementation. Gao et al introduced a compact fast elliptic curve crypto coprocessor with variable key size, which highly utilizes the internal *SRAM* and registers in a *Xilinx FPGA* [57].

In [58] the authors discuss reconfigurable *PAM* implementation of *RSA*. The *PAM* implementation of *RSA* used many advanced techniques in accelerating the execution time of the algorithm; for example, Chinese remainders, star chains, Hensel´s odd division, carry-save representation, quotient pipelining and asynchronous carry completion adders. An *FPGA*-based implementation of *DEA* is reported in [59] with a complete performance study and evaluation. Kim et al presented various architectures (low hardware complexity and high-performance versions) of the *KASUMI 3GPP* block cipher using a *Xilinx FPGA* [60].

Adam et al studied in [51] the hardware implementation within commercially available *FPGAs* of the potential *AES* candidates. Multiple architectural implementation options were explored for each algorithm.

## 2.6. High-Level Reconfigurable Hardware Development

The need for higher level solutions has been always present when using any computer architecture. An example high-level solution was the historic leap from punched cards to assembly language that was made with the early advancement in computers. Another example is the move up to high-level programming languages with the support of many developed design methodologies and paradigms. Another step up has been made with the languages capable of automatic code generation.

*Verilog* and *VHDL* (Very High Speed Integrated Circuit Hardware Description Language) [61] are by far the most commonly used hardware description languages (*HDLs*) in industry. Both of these two *HDLs* support different styles for describing hardware, for example, behavioral style, structural gate-level style, etc. *VHDL* became an *IEEE* standard *1076* in 1987. *Verilog* became an *IEEE* standard *1364* in December 1995. The *Verilog* language uses the *module* construct to declare logic blocks (with several inputs and outputs). In *VHDL*, each structural block consists of an interface description and architecture. VHDL enables behavioral descriptions in Data flow and Algorithmic styles.

Focused efforts for creating tools with higher levels of abstraction lead to the production of many powerful modern hardware design tools. Ian Page and Wayne Luk developed a compiler that transformed a subset of *Occam* into a netlist [62]. Nearly ten years later we have seen the development of *Handel-C*, the first commercially available high-level language for targeting programmable logic devices. *Handel-C* is a parallel programming language based on the theories of communicating sequential processes (*CSP*) and *Occam* with a *C*-like syntax familiar to most programmers. This language is used for describing computations which are to be compiled into hardware [28].

Building on the work carried out in Oxford's Hardware Compilation Group by Page and Luk, Saul at Oxford's Programming Research Group introduced a different co-design compiler, *Dash FPGA*-Based Systems [63]. This compiler provides a co-synthesis and co-simulation environment for mixed *FPGA* and processor architectures. It compiles a *C*-like description to a solution containing both processors and custom hardware.

Luk and McKeever in [64] introduced *Pebble*, a simple language designed to improve the productivity and effectiveness of hardware design. This language improves productivity by adopting reusable word-level and bit-level descriptions which can be customized by different parameter values, such as design size and the number of pipeline stages. Such descriptions can be compiled without flattening into various *VHDL* dialects. *Pebble* improves design effectiveness by supporting optional constraint descriptions, such as placement attributes, at various levels of abstraction; it also supports runtime reconfigurable designs.

Todman and Luk in [65] proposed a method that combines declarative and imperative hardware descriptions. They investigated the use of *Cobble* language, which allows abstractions to be done in an imperative setting. Designs done in *Cobble* are to benefit from efficient bit-level implementations developed in *Pebble*. Transformations are suggested to allow the declarative *Pebble* blocks to be used in *Cobbles'* imperative programs.

Weinhardt in [66] proposes a high-level language programming approach for reconfigurable computers. This automatically partitions the design between hardware and software and synthesizes pipelined circuits from parallel *for* loops.

W. Najjar et al in [67] presented a high-level, algorithmic language and optimizing compiler for the development of image processing applications on *RC*-systems. *SA-C*, a single assignment variant of the *C* programming language, was designed for this purpose.

A prototype *HDL* called Lava is developed by Satnam Singh at *Xilinx* and Mary Sheeran and Koen Claessen at Chalmers University in Sweden [68]. Lava allows circuit tiles to be composed using powerful higher-order combinators. This language is embedded in the *Haskell* lazy functional programming language. *Xilinx* implementation of *Lava* is designed to support the rapid representation, implementation and analysis of high-performance *FPGA* circuits.

Besides the above advances in the area of high-level hardware synthesis, the current market has other tools employed to aid programmable hardware implementations. These tools include

*Forge* compiler from *Xilinx*, *SystemC* language, *Nimble* compiler for *Agileware* architecture from *Nimbel Technology*, and *Superlog*.

Other famous hardware design tools include *Altera's Quartus*, *Xilinx ISE*, *Mentor Graphics HDL Designer*, *Leonardo Spectrum*, *Precision Synthesis*, and *ModelSim*.

### 2.7. Future of RC-Systems at a Glance

Finally, after having this brief overview of reconfigurable computing world, we would like to stress the expectations attached to this field. Wim Roelandts, the president and CEO of *Xilinx*, said:" I expect that within the first ten years of the next millennium we will see programmable logic devices inside every piece of electronic equipment, because hardware will become just as programmable as software" [69]. A different point is made by Mangione-Smith et al in [70-73] setting expectations for the near future of reconfigurable computing, they said: "Mainstream microprocessor vendors will not adopt *FPGA* blocks into their products, and barring some dramatic discovery programmable logic will not be widely available on the main datapath of high performance processors. However, it does appear likely that *FPGA* blocks will soon be incorporated in embedded processor devices."

The trends in the development in that field could be summarized as the expectation of ever denser *FPGAs* with very low cost. Internet reconfigurable logic is of possible interest in designing hardware that can be reconfigured remotely. Besides, efforts could be invested in creating high-level, fully-integrated, development tools for both large and small designs. All in all, in the coming years, bigger, faster, and cheaper reusable programmable devices are likely to appear.

## 3. The MorphoSys

One of the emerging *RC* systems includes the *MorphoSys* designed and implemented at the University of California, Irvine. It has the block diagram shown in Figure 1. It is composed of the following components:

- An array of reconfigurable cells (*ReC*) called the *ReC* array.
- *ReC* array configuration data memory called context memory.
- A control processor (*TinyRISC*).
- A data buffer called the frame buffer.
- A *DMA* controller.

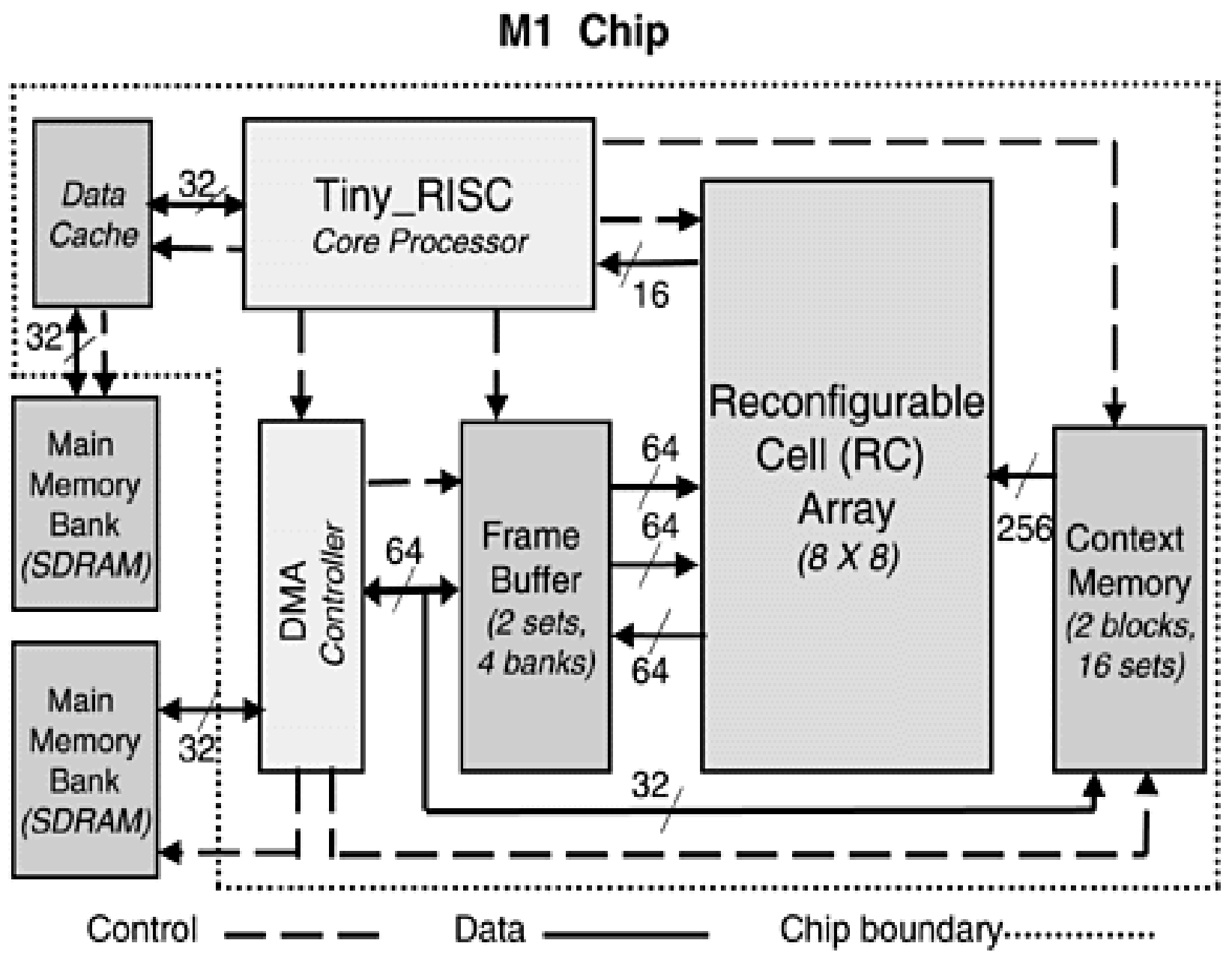

Figure 1. *MorphoSys* Block Diagram.

## 3.1. The Core Processor

The core processor known as *TinyRISC* is a *MIPS*-like processor with a stage scalar pipeline. The core processor has sixteen-bit registers, and three functional units; a bit *ALU*, a bit shift unit, and a memory unit. An on-chip data cache memory minimizes the accesses to external main memory. In addition to typical *RISC* instructions, *TinyRISC*s is augmented with specific instructions for controlling other *MorphoSys* components. The special instructions fall in two categories *DMA* instructions and *ReC* array instructions. *DMA* instructions initiate data transfers between main memory and the Frame Buffer and context loading from main memory into the Context Memory. *ReC* array instructions control the operation of the reconfigurable component the *ReC* array by specifying the context and the broadcast mode.

The reconfiguration instructions are classified into two sections *DMA* instructions and *ReC* array instructions. The *DMA* instructions specify load/store, memory address, number of bytes to be transferred, and the destination storage address (frame buffer or context memory). The *ReC* array instructions specify the context for execution, frame buffer address and broadcast mode (row, column, broadcast, or selective).

## 3.2. The Reconfigurable Cell

The *ReC* is the basic programmable element in *MorphoSys*. Each *ReC* comprises five components; the *ALU*-Multiplier, the shift unit, the input multiplexers, a register file with four bit registers, and the context register.

## 3.3. The Reconfigurable Cell Array

The reconfigurable part of the *MorphoSys* is called the *ReC* array that contain 64 *ReCs* arranged as an *8 x 8* matrix (See Figure 2). An important feature of the *ReC* array is its three-layer interconnection network, which enables two-dimensional mesh topology, complete row and column connectivity within a quadrant, and inter-quadrant connectivity.

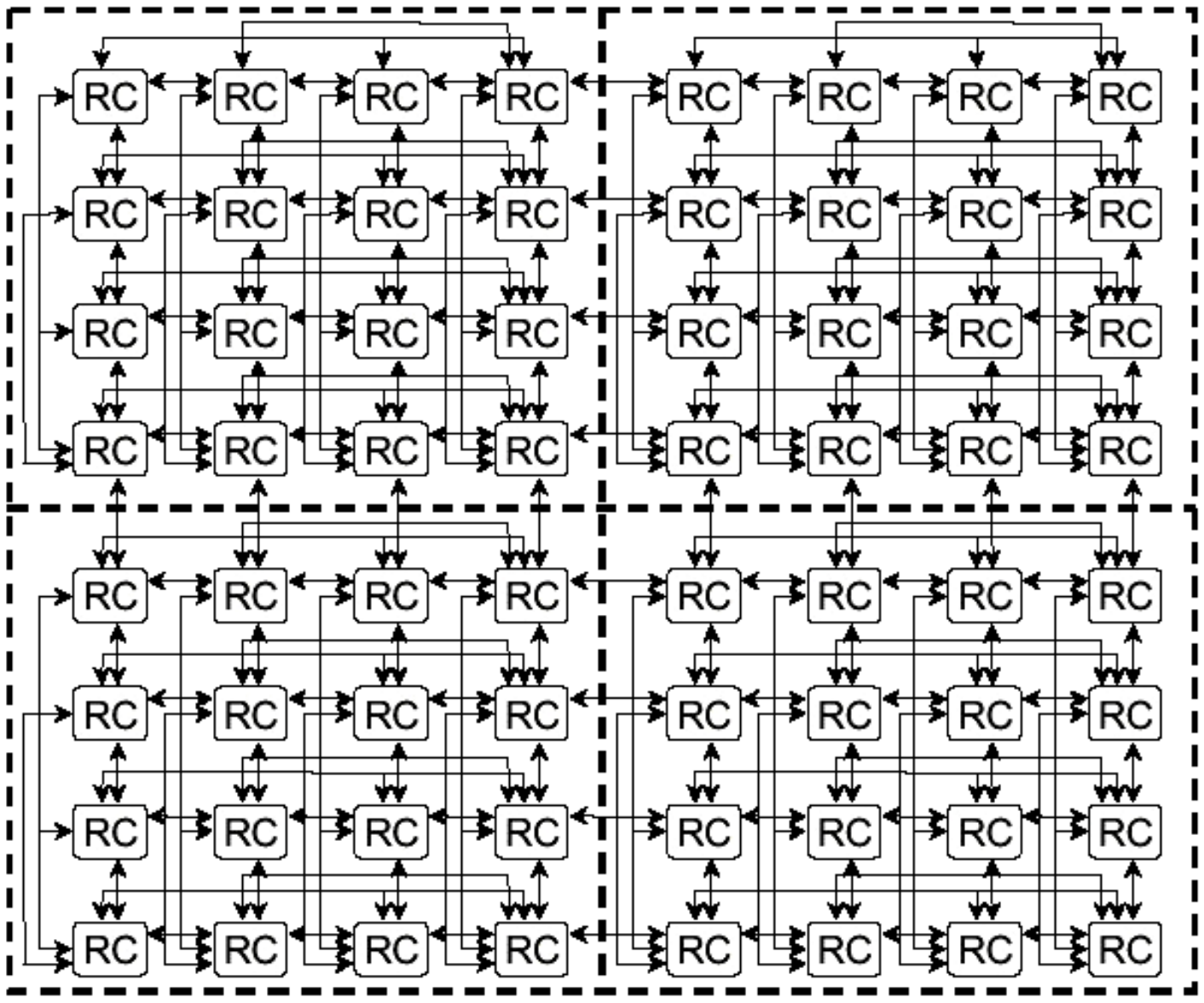

Figure 2. *ReC* Array Interconnection.

The *ReC* array operates in one of two modes; column context broadcast mode in which all the cells of a column perform the same operation, or row context broadcast mode in which all the cells of a row perform the same operation. Any application to be mapped onto *MorphoSys* and expected to make use of the reconfigurable core has to use the functions provided by the *ReCs* and specific interconnections that will enhance application performance by increasing the speed of execution of the application. Each *ReC* cell has two ports, *A* and *B*, through which it has access to: 1) the output of other cells, 2) the operand data bus, and 3) the internal register file of the *ReC*.

## 3.4. The Context Memory

The Context Memory stores the configuration program context for the *ReC* array. The Context Memory is logically organized into two context blocks each block containing eight context sets Each context set has sixteen context words. The major focus of the *ReC* array is on data-parallel applications, which exhibit a definite regularity. Following this principle of regularity and parallelism the context is broadcast on a row/column basis. The context words from one context memory block are broadcast along the rows, while context words from the other block are broadcast along the columns. Each block has eight context sets and each context set is associated with a specific row/column of the *ReC* array. The context word from a context set is broadcast to all eight *ReCs* in the corresponding row/column. Thus all *ReCs* in a row/column share a context word and perform the same operations. Recall that a context word is stored in the context register within each *ReC*.

A context plane is formed by the corresponding context words within each context set across the Context Memory. As there are sixteen context words in a context set up to sixteen context planes can be simultaneously resident in each of the two blocks of Context Memory.

## 3.5. The Frame Buffer and the DMA Controller

The Frame Buffer is an internal data memory logically organized into two sets. Each set is further subdivided into two banks. Each bank has 64 rows of  bytes (therefore, the entire Frame Buffer has 128 x 16 bytes). A 128-bit operand bus is used to transfer data operands from the Frame Buffer to the *ReC* array. This bus is connected to the column elements of the *ReC* array. The cells along a *ReC* array row share the same 16-bit segment of the operand bus. In this way, eight different operands can be loaded into all cells of a *ReC* array column in just a single cycle. Results from the *ReC* Array are written back to the Frame Buffer through a separate 128-bit bus called the result bus. The physical connection of the result bus to the *ReC* array is similar to that of the operand bus, i.e., the bus segments running along the rows.

The *DMA* controller performs data transfers between the Frame Buffer and the main memory. It is also responsible for loading contexts into the Context Memory. The *TinyRISC* core processor uses *DMA* instructions to specify the necessary data-context transfer parameters for the *DMA* controller.

## 3.6. The *MorphoSys* Execution Flow Model

A program runs on *MorphoSys* so that general-purpose operations are handled by the *TinyRISC* processor, while operations that have a certain degree of parallelism, regularity, or intensive computations are mapped to the *ReC* array. The *TinyRISC* processor controls, through the *DMA* controller, the loading of the context words to context memory. These context words define the function and connectivity of the cells in the *ReC* array. The processor also initiates the loading of application data, such as image frames, from main memory to the frame buffer.  This is also done through the *DMA* controller. Now that both configuration and application data are ready, the *TinyRISC* processor instructs the *ReC* array to start execution. The *ReC* array performs the needed operation on the application data and writes it back to the frame buffer. The *ReC* array loads new application data from the frame buffer and possibly new configuration data from context memory. Since the frame buffer is divided into two sets, new application data can be loaded into it without interrupting the operation of the *ReC* array.  Configuration data is also loaded into context memory without interrupting *ReC* array operation; this option allows the *MorphoSys* to expectedly achieve high speeds of execution.

## 3.7. Important Features of MorphoSys

*MorphoSys* is a coarse-grain multiple-context reconfigurable system with considerable depth of programmability (32 contexts) and two different context broadcast modes. It provides a high degree of flexibility for application mapping, by offering two levels of reconfigurability:

*Operand-type configurability:* this is configurability that allows switching between classes of applications with different data types. It affects two aspects of the system: (a) configuration of the array multiplier in each *ReC* as either a signed or unsigned multiplier, and (b) configuration of the operand bus (as either interleaved or contiguous) and of the result bus (as either 8-bit or 16-bit).

*Functional configurability:* this is the short-term, run-time reconfigurability level. It controls *ReC* functionality and *ReC* Array connectivity on a cycle-to-cycle basis.

The hierarchical *ReC* Array interconnection network also contributes for algorithm mapping flexibility Structures like the express lanes enhance global connectivity. Even irregular communication patterns, that otherwise require extensive interconnections can be handled efficiently. For instance, an eight-point butterfly can be accomplished in only three cycles. Finally, bus configurability supports applications with different data sizes and data flow patterns.

# 4. Graphics Geometrical Transformations under *MorphoSys*

Graphics hardware accelerators are of major importance to modern high-quality computer graphics. Geometrical transformations play a major role in computer graphics. Transformations, with their computational complexity, are usually supported in graphics accelerators. In this chapter, and for the purpose of hardware acceleration, parallel versions of basic and composite *2D* transformations are developed and implemented under *MorphoSys* [74].

## 4.1. Geometrical Transformations

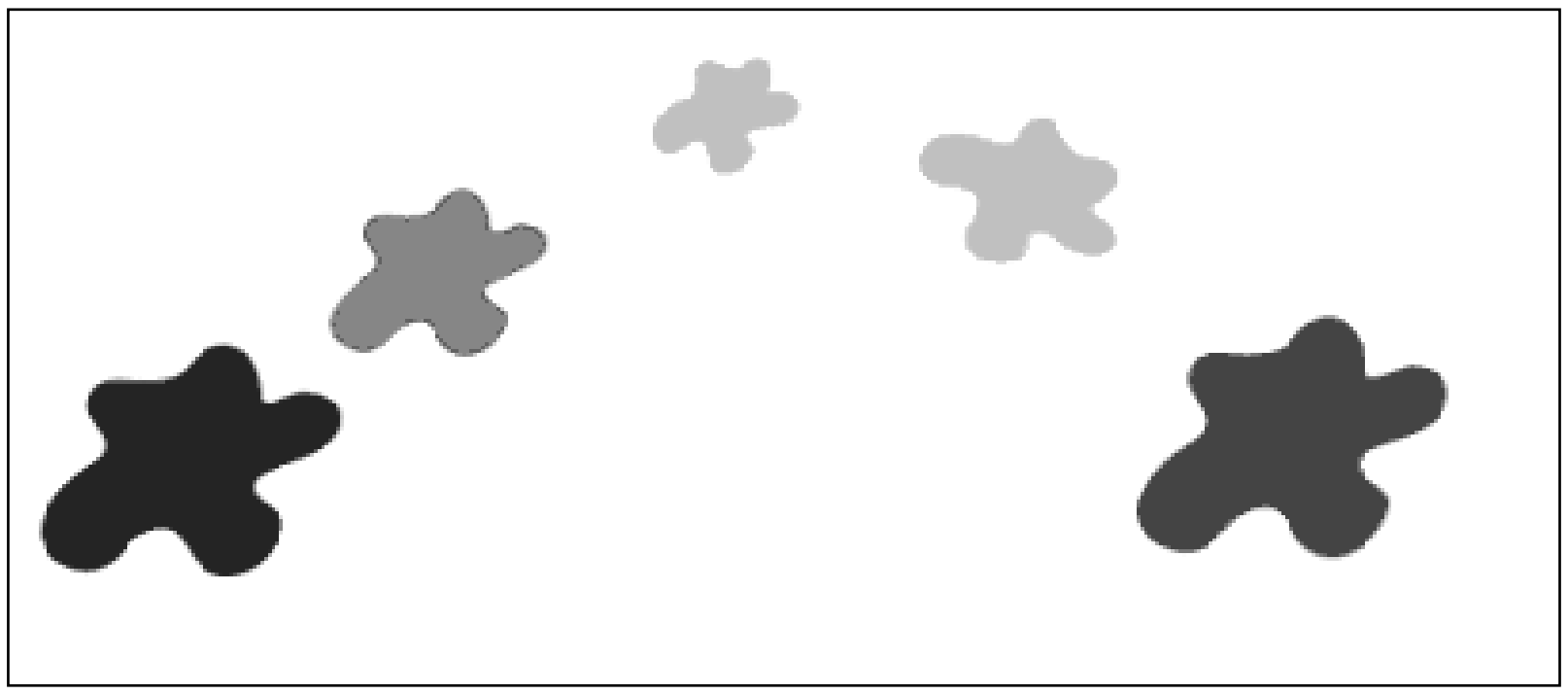

Figure 3. Image tracking while applying different *2D* transformations.

Transformations are a fundamental part of computer graphics. Transformations are used to position, shape, and change viewing positions of objects, as well as change how they are viewed (e.g. the type of projection used). Basic transformations include Translation, Scaling, Rotation, and Shear. These basic transformations can also be combined to obtain more complex transformations. Figure 3 shows the effects of some *2D* transformations on an image.

A point in a *2D* space could be represented by its name, abscissa, and ordinate; for instance, *P(x, y)* is a point with coordinates *x* and *y*. *2D* objects are often represented as a set of points (vertices), $\{P_1,P_2,...,P_n\}$, and an associated set of edges; $\{e_1,e_2,...,e_m\}$. An edge is defined as a pair of points, $e\{P_i, P_j\}$. We can also represent points in vector/matrix notation as:

$$\mathbf{P} = \begin{bmatrix} \mathbf{x} \\ \mathbf{y} \end{bmatrix}$$

### 4.1.1. Translations

A translation can also be represented by a pair of numbers, $t = (t_x, t_y)$, where $t_x$ is the change in the abscissa and $t_y$ is the change in the ordinate. To translate the point *P* by *t*, we simply add to obtain the translated point *Q(x',y')*.

$$Q = \begin{bmatrix} \mathbf{x} \\ \mathbf{y} \end{bmatrix} + \begin{bmatrix} \mathbf{t_x} \\ \mathbf{t_y} \end{bmatrix} = \begin{bmatrix} \mathbf{x} + \mathbf{t_x} \\ \mathbf{y} + \mathbf{t_y} \end{bmatrix}$$

In order to make translation into a multiplication homogeneous coordinates could be used [71]. Then, the matrix form for the translation is:

$$\begin{bmatrix} x' \\ y' \\ 1 \end{bmatrix} = \begin{bmatrix} 1 & 0 & t_x \\ 0 & 1 & t_y \\ 0 & 0 & 1 \end{bmatrix} \begin{bmatrix} x \\ y \\ 1 \end{bmatrix}$$

Homogeneous coordinates allow many transformations to be concatenated into a single matrix representation.

### 4.1.2. Scaling

Matrices can easily represent scaling transformations. Let the scale matrix be *S*, where:

$$S = \begin{bmatrix} \mathbf{s_x} & 0 \\ \mathbf{0} & \mathbf{s_y} \end{bmatrix}$$

The scaled point *Q(x',y')* could be determined as follows:

$$Q = S.P = \begin{bmatrix} S_x & 0 \\ 0 & S_y \end{bmatrix} \times \begin{bmatrix} x \\ y \end{bmatrix} = \begin{bmatrix} xS_\mathbf{x} \\ yS_y \end{bmatrix}$$

The matrix form for scaling using homogeneous coordinates is as follows:

$$\begin{bmatrix} x' \\ y' \\ 1 \end{bmatrix} = \begin{bmatrix} S_x & 0 & 0 \\ 0 & S_y & 0 \\ 0 & 0 & 1 \end{bmatrix} \begin{bmatrix} x \\ y \\ 1 \end{bmatrix}$$

The matrix product is also called compounding, catenation, concatenation or composition. Once the transform is formulated, it may be applied to all points in the scene. All this applies to other transformations also. Homogeneous coordinates allow many scaling operations to be concatenated into a single matrix representation.

### 4.1.3. Rotation and Shearing

For rotation by an angle $\theta$ counterclockwise about the origin, the functional form is $x' = x\cos\theta - y\sin\theta$ and $y' = x\sin\theta + y\cos\theta$. Written in matrix form, this becomes as follows:

$$Q = R.P = \begin{bmatrix} \cos\vartheta & -\sin\vartheta \\ \sin\vartheta & \cos\vartheta \end{bmatrix} \times \begin{bmatrix} x \\ y \end{bmatrix} = \begin{bmatrix} x\cos\vartheta - y\sin\vartheta \\ x\sin\vartheta + y\cos\vartheta \end{bmatrix}$$

The matrix form for scaling using homogeneous coordinates is as follows:

$$\begin{bmatrix} x' \\ y' \\ 1 \end{bmatrix} = \begin{bmatrix} \cos\vartheta & -\sin\vartheta & 0 \\ \sin\vartheta & \cos\vartheta & 0 \\ 0 & 0 & 1 \end{bmatrix} \begin{bmatrix} x \\ y \\ 1 \end{bmatrix}$$

For shearing, there are two possibilities. A shear parallel to the $x$ axis has $x' = x + ky$ and $y' = y$; the matrix form is:

$$Q = H_x.P = \begin{bmatrix} 1 & k \\ 0 & 1 \end{bmatrix} \times \begin{bmatrix} x \\ y \end{bmatrix} = \begin{bmatrix} x + ky \\ y \end{bmatrix}$$

A shear parallel to the $y$ axis has $x' = x$ and $y' = y + kx$, which has matrix form:

$$Q = H_y.P = \begin{bmatrix} 1 & 0 \\ k & 1 \end{bmatrix} \times \begin{bmatrix} x \\ y \end{bmatrix} = \begin{bmatrix} x \\ y + kx \end{bmatrix}$$

The matrix form for shearing using homogeneous coordinates is as follows:

$$\begin{bmatrix} x' \\ y' \\ 1 \end{bmatrix} = \begin{bmatrix} 1 & k & 0 \\ 0 & 1 & 0 \\ 0 & 0 & 1 \end{bmatrix} \begin{bmatrix} x \\ y \\ 1 \end{bmatrix}$$

and

$$\begin{bmatrix} x' \\ y' \\ 1 \end{bmatrix} = \begin{bmatrix} 1 & 0 & 0 \\ k & 1 & 0 \\ 0 & 0 & 1 \end{bmatrix} \begin{bmatrix} x \\ y \\ 1 \end{bmatrix}$$

It is clear from the presented formulas that *2D* geometrical transformation, basic or composite, are at the end a computationally intensive matrix multiplication problem.

## 4.2. Mapping Translation and Scaling in Basic Forms

In geometrical transformations, points are represented by vectors. To scale or translate vectors they are operated on by adding (subtracting) or multiplying (dividing) with constant values (scalars) that also could be represented using vectors. In *MorphoSys* system, the *M1* chip version, some operations could be classified as vector-vector and vector-scalar operations. These operations are used to implement basic translation, scaling, composite translations, and composite scaling.

### 4.2.1. Translation Using Vector-Vector Operations

A one-dimensional n-element vector could have the following transposed form:

$$\mathbf{M^T = [M_0 M_1 M_2 .....M_{n-1}]}$$

A vector *U* could be considered as the original coordinates, while a vector *V* could be considered as the corresponding translation values. Mapping an algorithm for addition, or any other operation, of the two vectors is done by first storing them in the Frame Buffer set "*0*" and set "*1*". Then we can exploit the properties of the interconnection, where some contents of Frame Buffer set "*0*" are added to some contents of Frame Buffer set "*1*" and the result would be in columns *0-7* of the *ReC* array. Figure 4 shows the final output in the *ReC* array after running the algorithm of adding two 64-element vectors.

| **Columns \| Rows** | **$C_0$** | **$C_1$** | **$C_2$** | **$C_3$** | **$C_4$** | **$C_5$** | **$C_6$** | **$C_7$** |
|---|---|---|---|---|---|---|---|---|
| **$R_0$** | $U_0+V_0$ | $U_8+V_8$ | $U_{16}+V_{16}$ | $U_{24}+V_{24}$ | $U_{32}+V_{32}$ | $U_{40}+V_{40}$ | $U_{48}+V_{48}$ | $U_{56}+V_{56}$ |
| **$R_1$** | $U_1+V_1$ | $U_9+V_9$ | $U_{17}+V_{17}$ | $U_{25}+V_{25}$ | $U_{33}+V_{33}$ | $U_{41}+V_{41}$ | $U_{49}+V_{49}$ | $U_{57}+V_{57}$ |
| **$R_2$** | $U_2+V_2$ | $U_{10}+V_{10}$ | $U_{18}+V_{18}$ | $U_{26}+V_{26}$ | $U_{34}+V_{34}$ | $U_{42}+V_{42}$ | $U_{50}+V_{50}$ | $U_{58}+V_{58}$ |
| **$R_3$** | $U_3+V_3$ | $U_{11}+V_{11}$ | $U_{19}+V_{19}$ | $U_{27}+V_{27}$ | $U_{35}+V_{35}$ | $U_{43}+V_{43}$ | $U_{51}+V_{51}$ | $U_{59}+V_{59}$ |
| **$R_4$** | $U_4+V_4$ | $U_{12}+V_{12}$ | $U_{20}+V_{20}$ | $U_{28}+V_{28}$ | $U_{36}+V_{36}$ | $U_{44}+V_{44}$ | $U_{52}+V_{52}$ | $U_{60}+V_{60}$ |
| **$R_5$** | $U_5+V_5$ | $U_{13}+V_{13}$ | $U_{21}+V_{21}$ | $U_{29}+V_{29}$ | $U_{37}+V_{37}$ | $U_{45}+V_{45}$ | $U_{53}+V_{53}$ | $U_{61}+V_{61}$ |
| **$R_6$** | $U_6+V_6$ | $U_{14}+V_{14}$ | $U_{22}+V_{22}$ | $U_{30}+V_{30}$ | $U_{38}+V_{38}$ | $U_{46}+V_{46}$ | $U_{54}+V_{54}$ | $U_{62}+V_{62}$ |
| **$R_7$** | $U_7+V_7$ | $U_{15}+V_{15}$ | $U_{23}+V_{23}$ | $U_{31}+V_{31}$ | $U_{39}+V_{39}$ | $U_{47}+V_{47}$ | $U_{55}+V_{55}$ | $U_{63}+V_{63}$ |

Figure 4. *ReC* array contents after vector addition.

For the *MorphoSys* to perform the required calculations, three sets of data must be first entered to the *M1* chip. The first data set is the *TinyRISC* program which is placed in main memory. The *TinyRISC* program handles all the operations that are not mapped onto the *ReC* array, such as, data transfer and *ReC* array contexts. The second data set contain the context codes of the *ReC* array. The context codes are written in either column mode, row mode, or for both. The context codes also define what operation each row or column is going to carry out, what input it takes, and where the output is to be stored. The last set entered to the *MorphoSys* is the input data set required for computations.

For translating vectors, the desired function of the *ReC* array interconnection is $Out = A + B$. Assume that vector *U* is stored at address $10{,}000_{hex}$ of main memory, vector *V* stored at address $20{,}000_{hex}$, and the context word stored at address $30{,}000_{hex}$. The output is stored back at address $40{,}000_{hex}$. The *TinyRISC* code for translating a 64-element vector with its discussion is provided in Table 1.

**Table 1. *TinyRISC* code translating a 64-element vector, (a) addresses setup, (b) broadcast data and operations, (c) write-back output.**

| | | | |
|---|---|---|---|
| 0: | ldui | r1, 0x1; | R1 ← $10000_{hex}$. Vector U is stored. |
| 1: | ldfb | r1, 0, 0, 16 ; | FB ← 16 x 32 bits at set 0, bank A, address 0. |
| 2: | add | r0, r0, r0; | No-operation. |
| . | . | . | |
| 33: | ldui | r1, 0x2; | R1 ← $20000_{hex}$.<br>This is where vector V is stored. |
| 34: | ldfb | r1, 1, 0, 16; | FB ← 16 x 32 bits at set 0, bank B, address 0. |
| 35: | add | r0, r0, r0; | NOP |
| . | . | | |
| 66: | ldui | r3, 0x3; | R3 ← $30000_{hex}$. This is where the context word is stored in main memory. |
| 67: | ldctxt | r3, 0, 0, 0, 1; | Load one context word from main memory starting at the address stored in register 3 into plane 0, block 0 and starting at word 0. |
| 68: | add | r0, r0, r0; | NOP |
| . | . | . | |

**(a)**

| | | | |
|---|---|---|---|
| 71: | ldui | r4, 0x0; | R4 ← $00000_{hex}$. |
| 72: | dbcdc | r4, 0, 0, 0, 0, 0, 0; | Double bank column broadcast. It sends data from both banks address 0 in the frame buffer and broadcasts the context words column-wise. It triggers the RC array to start execution of column 0 by the context word of address 0 in the column block of context memory operating on data in set 0. Bank A starting at 0x0. Bank B starting at (0x0 + 0). |
| 73: | ldli | r4, 0x4 | R4 ← $4_{hex}$ |

**Table 1. Continued.**

| | | | |
|---|---|---|---|
| 74: | dbcdc | r4, 0, 0, 1, 0, 0, 0x40; | It sends data from both banks address $40_{hex}$ in the frame buffer. Bank A starting at 0x40. Bank B starting at (0x4 + 0x0 = 0x40). |
| 75: | ldli | r4, 0x8 | R4 ← $8_{hex}$ |
| 76: | dbcdc | r4, 0, 0, 2, 0, 0, 0x80; | It sends data from both banks. |
| 77: | ldli | r4, 0xC | R4 ← $C_{hex}$ |
| 78: | dbcdc | r4, 0, 0, 3, 0, 0, 0xC0; | It sends data from both banks address $C0_{hex}$ in the frame buffer. Bank A starting at 0xC0. Bank B starting at (0xC + 0x0 = 0xC0). |
| 79: | ldli | r4, 0x10 | R4 ← $10_{hex}$ |
| 80: | dbcdc | r4, 0, 0, 4, 0, 0, 0x100; | It sends data from both banks address $100_{hex}$ in the frame buffer. Bank A starting at 0x100. Bank B starting at (0x10 + 0x0 = 0x100). |
| 81: | ldli | r4, 0x14 | R4 ← $14_{hex}$ |
| 82: | dbcdc | r4, 0, 0, 5, 0, 0, 0x140; | It sends data from both banks address $140_{hex}$ in the frame buffer. Bank A starting at 0x140. Bank B starting at (0x14 + 0x0 = 0x140). |
| 83: | ldli | r4, 0x18 | R4 ← $18_{hex}$ |
| 84: | dbcdc | r4, 0, 0, 6, 0, 0, 0x180; | It sends data from both banks. |
| 85: | ldli | r4, 0x1C | R4 ← $1C_{hex}$ |
| 86: | dbcdc | r4, 0, 0, 7, 0, 0, 0x1C0; | It sends data from both banks. |

**(b)**

| | | | |
|---|---|---|---|
| 87: | wfbi | 0, 0, 0, 1, 0x0; | Write data back to the frame buffer from the output registers of column 0 into set 1, address 0. |
| 88: | wfbi | 1, 0, 0, 1, 0x40; | of column 1 into set 1, address 64. |
| 89: | wfbi | 2, 0, 0, 1, 0x80; | of column 2 into set 1, address 128. |
| 90: | wfbi | 3, 0, 0, 1, 0xC0; | of column 3 into set 1, address 192. |
| 91: | wfbi | 4, 0, 0, 1, 0x100; | of column 4 into set 1, address 256. |
| 92: | wfbi | 5, 0, 0, 1, 0x140; | of column 5 into set 1, address 320. |
| 93: | wfbi | 6, 0, 0, 1, 0x180; | of column 6 into set 1, address 384. |
| 94: | wfbi | 7, 0, 0, 1, 0x1C0; | of column 7 into set 1, address 448. |
| 95: | ldui | r5, 0x4; | R5 ← 40000hex. |
| 96: | stfb | r1, 1, 0,$10_{hex}$; | Store data from frame buffer set 1, address 0 into main memory starting at address stored in reg1. |

**(c)**

### 4.2.2. Scaling Using Vector-Scalar Operations

Consider the following 8-element vector *U*:

$$\mathbf{U^T = [U_0 \quad U_1 \quad U_2 \quad U_3 \quad U_4 \quad U_5 \quad U_6 \quad U_7]}$$

Scaling *U* with a constant *c*, the following is obtained:

$$\mathbf{W = c \times U \Rightarrow \quad W^T = [cU_0\ cU_1\ ....cU_7]}$$

For *n*-element vectors:

$$\mathbf{W^T = [cU_0\ cU_1.....cU_n]}.$$

Mapping the algorithm for multiplication, or any other operation (arithmetic or logical), of a vector by a scalar, is done by first storing the vector in the Frame Buffer set "*0*". Then we can exploit the properties of the interconnection, where some contents of the Frame Buffer set "*0*" are multiplied by a constant to be stored in the context word. Figure 8 shows the final output in the *ReC* array after running the algorithms of two 64-element vectors.

The desired function of the interconnection is: *Out (t+1) = c x A*. Assume that a vector *U* is stored at address $30{,}000_{hex}$ of main memory, and the context word stored at address $40{,}000_{hex}$. Then the answer will be stored back at address $50{,}000_{hex}$. The code and its discussion for scaling a 64-element vector are provided in Table 2.

| **Columns\| Rows** | **$C_0$** | **$C_1$** | **$C_2$** | **$C_3$** | **$C_4$** | **$C_5$** | **$C_6$** | **$C_7$** |
|---|---|---|---|---|---|---|---|---|
| **$R_0$** | $c.U_0$ | $c.U_8$ | $c.U_{16}$ | $c.U_{24}$ | $c.U_{32}$ | $c.U_{40}$ | $c.U_{48}$ | $c.U_{56}$ |
| **$R_1$** | $c.U_1$ | $c.U_9$ | $c.U_{17}$ | $c.U_{25}$ | $c.U_{33}$ | $c.U_{41}$ | $c.U_{49}$ | $c.U_{57}$ |
| **$R_2$** | $c.U_2$ | $c.U_{10}$ | $c.U_{18}$ | $c.U_{26}$ | $c.U_{34}$ | $c.U_{42}$ | $c.U_{50}$ | $c.U_{58}$ |
| **$R_3$** | $c.U_3$ | $c.U_{11}$ | $c.U_{19}$ | $c.U_{27}$ | $c.U_{35}$ | $c.U_{43}$ | $c.U_{51}$ | $c.U_{59}$ |
| **$R_4$** | $c.U_4$ | $c.U_{12}$ | $c.U_{20}$ | $c.U_{28}$ | $c.U_{36}$ | $c.U_{44}$ | $c.U_{52}$ | $c.U_{60}$ |
| **$R_5$** | $c.U_5$ | $c.U_{13}$ | $c.U_{21}$ | $c.U_{29}$ | $c.U_{37}$ | $c.U_{45}$ | $c.U_{53}$ | $c.U_{61}$ |
| **$R_6$** | $c.U_6$ | $c.U_{14}$ | $c.U_{22}$ | $c.U_{30}$ | $c.U_{38}$ | $c.U_{46}$ | $c.U_{54}$ | $c.U_{62}$ |
| **$R_7$** | $c.U_7$ | $c.U_{15}$ | $c.U_{23}$ | $c.U_{31}$ | $c.U_{39}$ | $c.U_{47}$ | $c.U_{55}$ | $c.U_{63}$ |

Figure 5. *ReC* array contents after multiplication with a constant.

## 4.3. Basic Transformation Compositions

### 4.3.1. Composition of Two Translations

The composition of two successive translations could be viewed as a combination of both vector-vector and vector-scalar *MorphoSys* operations. Two successive translations means first translating a vector *V1* with translating values found in another vector *V2*. Then, exploiting the *MorphoSys ALU*-Features to perform the second translation by adding the result of the first translation to a third vector *V3* containing the second translating values. This second step is done by using a new vector-scalar operation, which suggests the addition of the output of the *ReC* array cells to another vector in the frame buffer. In other words, the algorithm could be divided into two steps:

**Table 2. *TinyRISC* code for the uniform scaling routine of a 64-element vector.**

| | | |
|---|---|---|
| 0: | Ldui r1, 0x1; | R1 ← $10000_{hex}$. This is where vector U is stored. |
| 1: | Ldfb r1, 0, 0, 16 ; | FB ← 16 x 32 bits at set 0, bank A, address 0. |
| 2: | Add r0, r0, r0; | No-operation. |
| . | . . | |
| 33: | ldui r3, 0x3; | R3 ← $30000_{hex}$. This is where the context word is stored in main memory. |
| 34: | ldctxt r3, 0, 0, 0, 1; | Load one context word from main memory starting at the address stored in register 3 into plane 0, block 0 and starting at word 0. |
| 35: | add r0, r0, r0; | NOP |
| . | . . | |
| 38: | sbcb 1, 0, 0, 0, 0, 0, 0x0; | Single bank column broadcast causing all the cells in the RC array to perform their operations specified by the context word in context memory starting with data from frame buffer, set 0, bank A, address offset 0. |
| 39: | sbcb 1, 0, 0, 0, 0, 0, 0x40; | It sends data from both banks address $40_{hex}$ in FB. |
| 40: | sbcb 1, 0, 0, 0, 0, 0, 0x80; | It sends data from both banks address $80_{hex}$ in FB. |
| 41: | sbcb 1, 0, 0, 0, 0, 0, 0xC0; | It sends data from both banks address $C0_{hex}$ in FB |
| 42: | sbcb 1, 0, 0, 0, 0, 0, 0x100; | It sends data from both banks address $100_{hex}$ in FB |
| 43: | sbcb 1, 0, 0, 0, 0, 0, 0x140; | It sends data from both banks address $140_{hex}$ in FB |
| 44: | sbcb 1, 0, 0, 0, 0, 0, 0x180; | It sends data from both banks address $180_{hex}$ in FB |
| 45: | sbcb 1, 0, 0, 0, 0, 0, 0x1C0; | It sends data from both banks address $1C0_{hex}$ in FB |
| 46: | wfbi 0, 0, 0, 1, 0x0; | Write data back to the frame buffer from the output registers<br>of column 0 into set 1, address 0. |
| 47: | wfbi 1, 0, 0, 1, 0x40; | of column 1 into set 1, address 64. |
| 48: | wfbi 2, 0, 0, 1, 0x80; | of column 2 into set 1, address 128. |
| 49: | wfbi 3, 0, 0, 1, 0xC0; | of column 3 into set 1, address 192. |
| 50: | wfbi 4, 0, 0, 1, 0x100; | of column 4 into set 1, address 256. |
| 51: | wfbi 5, 0, 0, 1, 0x140; | of column 5 into set 1, address 320. |
| 52: | wfbi 6, 0, 0, 1, 0x180; | of column 6 into set 1, address 384. |
| 53: | wfbi 7, 0, 0, 1, 0x1C0; | of column 7 into set 1, address 448. |
| 54: | ldui r5, 0x4; | R5 ← 40000hex. |
| 55: | stfb r1, 1, 0, $10_{hex}$; | Store data from frame buffer set 1, address 0 into main memory starting at address stored in reg1. |

The output of the *ReC* array is: *Out (t) = V1 + V2*, where *V1* and *V2* are found in the frame buffer.

The output is then added to the third vector *V3* using the context word that yields: *Out (t+1) = CxV3 +Out (t)*, where *C* is considered to be a constant of unity value and *V3* a vector in the frame buffer. Therefore, *Out (t+1) = V3 +Out (t)* or *Out (t+1) = V1 + V2 +V3*.

The mapping works as following: the contents of the Frame Buffer set "*0*" are added to the some contents of the Frame Buffer set "*1*" and the result would be in column *0*. This is demonstrated in Figure 6. The main motivation behind the composite translations mapping is it's expected major saving in cycles count over the composition using matrix operations.

| **Columns\| Rows** | **$C_0$** | **$C_1$** | **$C_2$** | **$C_3$** | **$C_4$** | **$C_5$** | **$C_6$** | **$C_7$** |
|---|---|---|---|---|---|---|---|---|
| **$R_0$** | $U_0 + V_0$ | | | | | | | |
| **$R_1$** | $U_1 + V_1$ | | | | | | | |
| **$R_2$** | $U_2 + V_2$ | | | | | | | |
| **$R_3$** | $U_3 + V_3$ | | | | | | | |
| **$R_4$** | $U_4 + V_4$ | | | | | | | |
| **$R_5$** | $U_5 + V_5$ | | | | | | | |
| **$R_6$** | $U_6 + V_6$ | | | | | | | |
| **$R_7$** | $U_7 + V_7$ | | | | | | | |

Figure 6. *ReC* array contents after addition is performed.

The next step is to take the output of the previous cycle and to change the *ReC*-interconnection and the *ALU*-operation to perform the next calculation (See Figure 7).

| **Columns\| Rows** | **$C_0$** | **…** | **$C_6$** | **$C_7$** |
|---|---|---|---|---|
| **$R_0$** | $c.U_0$ + out(t) | | | |
| **$R_1$** | $c.U_1$+ out(t) | | | |
| **$R_2$** | $c.U_2$+ out(t) | | | |
| **$R_3$** | $c.U_3$+ out(t) | | | |
| **$R_4$** | $c.U_4$+ out(t) | | | |
| **$R_5$** | $c.U_5$+ out(t) | | | |
| **$R_6$** | $c.U_6$+ out(t) | | | |
| **$R_7$** | $c.U_7$+ out(t) | | | |

Figure 7. The result of a scalar*vector operation for an *8*-element vector to a previously calculated output.

The vectors *V1*, *V2*, and *V3* could to be loaded into: *Bank A/Set 0*, *Bank B/Set 0*, *Bank A/Set 1* of the frame buffer and the constant is loaded in the context word. Assuming that the needed context word is available in main memory and that the vector is also available in main memory, the code for the *TinyRISC* has to take care of all the rest. The *TinyRISC* code has to load the vectors from main memory to the frame buffer. After that it has to load the needed context from main memory into the column block in context memory twice. Assume that vector *V1* is stored in address $10,000_{hex}$ of main memory, vector *V2* is stored in address $20,000_{hex}$, and vector *V3* is stored in address *30,000<sub></sub>*$_{hex}$. The context words are stored in address *40,000*$_{hex}$ and *50,000*$_{hex}$. The code and its discussion are in Table 3.

**Table 3. *TinyRISC* code for the composite translations routine (a) addresses setup, (b) broadcast data and operations, (c) write-back output.**

| | | | |
|---|---|---|---|
| 0: | ldui | r1, 0x1; | R1 ← 10000$_{hex}$. This is where vector $V_1$ is stored. |
| 1: | ldfb | r1, 0, 0, 16 ; | FB ← 16 x 32 bits at set 0, bank A, address 0. |
| 2: | add | r0, r0, r0; | No-operation. |
| . | . | . | |
| 33: | ldui | r1, 0x2; | R1 ← 20000$_{hex}$. This is where vector $V_2$ is stored. |
| 34: | ldfb | r1, 1, 0, 16 ; | FB ← 16 x 32 bits at set 0, bank B, and address 0. From main memory starting at the address stored in register 1. |
| 35: | add | r0, r0, r0; | No-operation. |
| . | . | . | |
| 66: | ldui | r1, 0x3; | R1 ← 30000$_{hex}$. This is where vector $V_3$ is stored. |
| 67: | ldfb | r1, 0, 1, 16; | FB ← 16 x 32 bits at set 1, bank A, address 0. |
| 68: | add | r0, r0, r0; | NOP |
| . | . | | |
| 99: | ldui | r3, 0x4; | R3 ← 40000$_{hex}$. This is where the context stored in memory. |
| 100: | ldctxt | r3, 0, 0, 0, 1; | Load one context word from main memory starting at the address stored in register 3 into plane 0, block 0 and starting at word 0. |
| 101: | add | r0, r0, r0; | NOP |
| . | . | . | |
| 104: | ldui | r3, 0x5; | R3 ← 50000$_{hex}$. This is where the context is stored in memory. |
| 105: | ldctxt | r3, 0, 0, 1, 1; | Load one context word from main memory starting at the address stored in register 3 into plane 1, block 0 and starting at word 0. |
| 106: | add | r0, r0, r0; | NOP |
| . | . | . | . |

**(a)**

| | | | |
|---|---|---|---|
| 109: | ldui | r4, 0x0; | R4 ← 00000$_{hex}$. |
| 110: | dbcdc | r4, 0, 0, 0, 0, 0, 0; | Double bank column broadcast. It sends data from both banks address 0 in the frame buffer and broadcasts the context words column-wise. It triggers the RC array to start execution of column 0 by the context word of address 0 in the column block of context memory operating on data in set 0. Bank A starting at 0x0. Bank B starting at (0x0 + 0). |
| 111: | ldli | r4, 0x4 | R4 ← 4$_{hex}$ |
| 112: | dbcdc | r4, 0, 0, 1, 0, 0, 0x40; | It sends data from both banks address 40$_{hex}$ in the frame buffer. Bank A starting at 0x40. Bank B starting at (0x4 + 0x0 = 0x40). |
| 113: | ldli | r4, 0x8 | R4 ← 8$_{hex}$ |

**Table 3. Continued.**

| | | | |
|---|---|---|---|
| 114: | Dbcd c | r4, 0, 0, 2, 0, 0, 0x80; | It sends data from both banks address $80_{hex}$ in the frame buffer. Bank A starting at 0x80. Bank B starting at (0x8 + 0x0 = 0x80). |
| 115: | ldli | r4, 0xC | R4 ← $C_{hex}$ |
| 116: | dbcdc | r4, 0, 0, 3, 0, 0, 0xC0; | It sends data from both banks address $C0_{hex}$ in the frame buffer. Bank A starting at 0xC0. Bank B starting at (0xC + 0x0 = 0xC0). |
| 117: | ldli | r4, 0x10 | R4 ← $10_{hex}$ |
| 118: | dbcdc | r4, 0, 0, 4, 0, 0, 0x100; | It sends data from both banks address $100_{hex}$ in the frame buffer. Bank A starting at 0x100. Bank B starting at (0x10 + 0x0 = 0x100). |
| 119: | ldli | r4, 0x14 | R4 ← $14_{hex}$ |
| 120: | dbcdc | r4, 0, 0, 5, 0, 0, 0x140; | It sends data from both banks address $140_{hex}$ in the frame buffer. Bank A starting at 0x140. Bank B starting at (0x14 + 0x0 = 0x140). |
| 121: | ldli | r4, 0x18 | R4 ← $18_{hex}$ |
| 122: | dbcdc | r4, 0, 0, 6, 0, 0, 0x180; | It sends data from both banks address $180_{hex}$ in the frame buffer. Bank A starting at 0x180. Bank B starting at (0x18 + 0x0 = 0x180). |
| 123: | ldli | r4, 0x1C | R4 ← $1C_{hex}$ |
| 124: | dbcdc | r4, 0, 0, 7, 0, 0, 0x1C0; | It sends data from both banks address $1C0_{hex}$ in the frame buffer. Bank A starting at 0x1C0. Bank B starting at (0x1C + 0x0 = 0x1C0). |
| 125: | sbcb | 1, 0, 0, 1, 0, 1, 0x0; | Single bank column broadcast causing all the cells in the RC array to perform their operations specified by the context word in context memory/plane, 1 starting with data from frame buffer, set 1, bank A, address offset 0. |
| 126: | sbcb | 1, 0, 0, 1, 0, 1, 0x40; | It sends data from bank A/Set 1 address 40hex in FB. |
| 127: | sbcb | 1, 0, 0, 1, 0, 1, 0x80; | It sends data from bank A/Set 1 address 80hex in FB. |
| 128: | sbcb | 1, 0, 0, 1, 0, 1, 0xC0; | It sends data from bank A/Set 1 address C0hex in FB |
| 129: | sbcb | 1, 0, 0, 1, 0, 1, 0x100; | It sends data from bank A/Set 1 address 100hex in the frame buffer. |
| 130: | sbcb | 1, 0, 0, 1, 0, 1, 0x140; | It sends data from bank A/Set 1 address 140hex in the buffer. |
| 140: | sbcb | 1, 0, 0, 1, 0, 1, 0x180; | It sends data from bank A/Set 1 address 180hex in the buffer. |
| 150: | sbcb | 1, 0, 0, 1, 0, 1, 0x1C0; | It sends data from bank A/Set 1 address 1C0hex in the buffer. |

**(b)**

**Table 3. Continued.**

| | | | |
|---|---|---|---|
| | wfbi | 0, 0, 0, 1, 0x0; | Write data back to the frame buffer from the output registers of column 0 into set 1, address 0. |
| 152: | wfbi | 1, 0, 0, 1, 0x40; | of column 1 into set 1, address 64. |
| 153: | wfbi | 2, 0, 0, 1, 0x80; | of column 2 into set 1, address 128. |
| 154: | wfbi | 3, 0, 0, 1, 0xC0; | of column 3 into set 1, address 192. |
| 155: | Wfbi | 4, 0, 0, 1, 0x100; | of column 4 into set 1, address 256. |
| 156: | wfbi | 5, 0, 0, 1, 0x140; | of column 5 into set 1, address 320. |
| 157: | wfbi | 6, 0, 0, 1, 0x180; | of column 6 into set 1, address 384. |
| 158: | wfbi | 7, 0, 0, 1, 0x1C0; | of column 7 into set 1, address 448. |
| 159: | ldui | r5, 0x4; | R5 ← 40000hex. |
| 160: | stfb | r1, 1, 0,$10_{hex}$; | Store data from frame buffer set 1, address 0 into main memory starting at address stored in register 1. |
| 161: | add | r0, r0, r0; | NOP |
| . | . | . | . |
| 224: | add | r0, r0, r0; | NOP |

**(c)**

### 4.3.2. Composition of Two Scaling Transformations

The composition of two scaling operations would mean either to calculate a composite scaling parameter, or to perform two successive scaling operations i.e. to call twice the subroutine suggested in (scalar-vector multiplication routine). This is similar to the case of any other geometrical transformation. Generally, to construct a subroutine, for two successive transformations, is useful for code reusability, especially, if we could obtain a major saving in execution time due to the use of such a subroutine. To perform the first way of performing two scaling operations, a composite scaling parameter is calculated. This scaling parameter is the result of multiplying the two scaling factors. Then, to perform the scaling of the points, which is also the multiplication of the scaled points with the composite scaling factor. Using the *MorphoSys ALU* basic operations the steps could be written as following:

$$Out(t) = C1xC2$$
$$Out(t+1) = Out(t) \ x \ A$$

In a different context:

$$Out(t) = C1xA.$$
$$Out(t+1) = C2 \ x \ Out(t).$$

Both the previous steps require two unsupported operations in the *ALU* of the *MorphoSys ReC* array. This led us to the possibility of modifying the *ALU* operations to support the operations $Out(t+1) = Out(t) \ x \ A$ (where $A$ is a vector), $Out(t+1) = C \ x \ Out(t)$ (where $C$ is a scalar), and $Out\ (t) = B \ x \ A$ (where $B$ is a vector).

Discussing only the first modification that assumes the presence of the operation *Out(t+1) = C x Out(t)*, this modification assumes the presence of the multiply operation in the *ReCs*. The code after this modification is discussed in Table 4.

### 4.3.3. Composition of Translation and Scaling

For a composite translation and scaling, the desired output at the *ReCs* is:

*Out(t) = (C x A) + B*, where *A* and *B* are vectors, while *C* is a constant.

The code for such a composite transformation is similar to what is presented in the previous code segments.

**Table 4. *TinyRISC* code, composite scaling operations using the suggested modification.**

| | | | |
|---|---|---|---|
| 0: | ldui | r1, 0x1; | R1 ← $10000_{hex}$. This is where vector A is stored. |
| 1: | ldfb | r1, 0, 0, 16 ; | FB ← 16 x 32 bits at set 0, bank A, address 0. |
| 2: | add | r0, r0, r0; | No-operation. |
| . | . | . | |
| 33: | ldui | r3, 0x4; | R3 ← $40000_{hex}$. where the context word is in mem. |
| 34: | ldctxt | r3, 0, 0, 0, 1; | Load one context word from main memory starting at the address stored in register 3 into plane 0, block 0 and starting at word 0. |
| 35: | add | r0, r0, r0; | NOP |
| . | . | . | |
| 38: | ldui | r3, 0x5; | R3 ← $50000_{hex}$. where the context word is in mem. |
| 39: | ldctxt | r3, 0, 0, 1, 1; | Load another context word from main memory starting at the address stored in register 3 into plane 1, block 0 and starting at word 0. |
| 40: | Add | r0, r0, r0; | NOP |
| . | . | . | . |
| 42: | ldui | r4, 0x0; | R4 ← $00000_{hex}$. |
| 43: | sbcb | 1, 0, 0, 0, 0, 0, 0x0; | Single bank column broadcast causing all the cells in the RC array to perform their operations specified by the context word in context memory starting with data from frame buffer, set 0, bank A, address offset 0. |
| 44: | sbcb | 1, 0, 0, 0, 0, 0, 0x40; | It sends data from bank at address $40_{hex}$ in FB. |
| 45: | sbcb | 1, 0, 0, 0, 0, 0, 0x80; | It sends data from bank at address $80_{hex}$ in FB. |
| 46: | sbcb | 1, 0, 0, 0, 0, 0, 0xC0; | It sends data from bank at address $C0_{hex}$ in FB |
| 47: | sbcb | 1, 0, 0, 0, 0, 0, 0x100; | It sends data from bank at address $100_{hex}$ in the FB. |
| 48: | sbcb | 1, 0, 0, 0, 0, 0, 0x140; | It sends data from bank at address $140_{hex}$ in the FB. |
| 49: | sbcb | 1, 0, 0, 0, 0, 0, 0x180; | It sends data from bank at address $180_{hex}$ in the FB. |
| 50: | sbcb | 1, 0, 0, 0, 0, 0, 0x1C0; | It sends data from bank address $1C0_{hex}$ in the FB . |

**Table 4. Continued.**

| | | |
|---|---|---|
| 51: | sbcb 1, 0, 0, 1, 0, 1, 0x0; | Single bank column broadcast causing all the cells in the RC array to perform their operations specified by the new suggested context word (with ALU enhancement) in context memory causing Out (t+1) = C2 x Out (t). Where, no data is taken from the frame buffer. |
| 52: | sbcb 1, 0, 0, 1, 0, 1, 0x0; | |
| 53: | sbcb 1, 0, 0, 1, 0, 1, 0x0; | |
| 54: | sbcb 1, 0, 0, 1, 0, 1, 0x0; | |
| 55: | sbcb 1, 0, 0, 1, 0, 1, 0x0; | |
| 56: | sbcb 1, 0, 0, 1, 0, 1, 0x0; | |
| 57: | sbcb 1, 0, 0, 1, 0, 1, 0x0; | |
| 58: | sbcb 1, 0, 0, 1, 0, 1, 0x0; | |
| 59: | Wfbi 0, 0, 0, 1, 0x0; | Write data back to the frame buffer from the output registers of column 0 into set 1, address 0. |
| 60: | wfbi 1, 0, 0, 1, 0x40; | of column 1 into set 1, address 64. |
| 61: | wfbi 2, 0, 0, 1, 0x80; | of column 2 into set 1, address 128. |
| 62: | wfbi 3, 0, 0, 1, 0xC0; | of column 3 into set 1, address 192. |
| 63: | wfbi 4, 0, 0, 1, 0x100; | of column 4 into set 1, address 256. |
| 64: | wfbi 5, 0, 0, 1, 0x140; | of column 5 into set 1, address 320. |
| 65: | wfbi 6, 0, 0, 1, 0x180; | of column 6 into set 1, address 384. |
| 66: | wfbi 7, 0, 0, 1, 0x1C0; | of column 7 into set 1, address 448. |

## 4.4. Transformations using the General Matrix Form

All basic and composite transformations, such as rotation and shearing, could be performed using the general matrix form. In this section, a parallel matrix multiplication algorithm is developed and mapped onto the *MorphoSys*. Because of the *ReC* array size, the matrices dimensions are chosen to be *8x8*.

In the standard matrix multiplication algorithm, multiplying a matrix *A* with a matrix *B* would mean multiplying row one ($r_1$) of *A* with column one of *B* and then adding their results yielding ($c_{11}$) of the resultant matrix *C*. The multiplication with ($r_1$) is repeated to all columns of *B* resulting in ($c_{12}$ ... $c_{1n}$). Then, ($r_2$) of *A* is multiplied with all columns of *B*. This algorithm is repeated till the last row in *A*. Matrices *A*, *B*, and *C* are dense matrices.

### 4.4.1. First Mapping

This simple algorithm could be mapped onto the *ReC a*rray as follows: The contents of the matrix *A* are passed row by row through the context words, thus, stored in the context memory for later retrieval and manipulation by the *ReCs*. The contents of matrix *B* are broadcasted also row by row to the columns of the *ReC* array (See Figures 8, 9 and 10). The multiplication stage (row x column) is done by using the *CMUL ALU* operation where *Out(t) = C x A*; which is the required computation. Note that *CMUL* is a vector-scalar operation.

| CMUL by | $a_{11}$ | $a_{12}$ | $a_{13}$ | $a_{14}$ | $a_{15}$ | $a_{16}$ | $a_{17}$ | $a_{18}$ |
|---|---|---|---|---|---|---|---|---|
| **Columns\|Rows** | **$C_0$** | **$C_1$** | **$C_2$** | **$C_3$** | **$C_4$** | **$C_5$** | **$C_6$** | **$C_7$** |
| **$R_0$** | $b_{11}$ | $b_{21}$ | . | . | . | . | . | . |
| **$R_1$** | $b_{12}$ | $b_{22}$ | . | . | . | . | . | . |
| **$R_2$** | $b_{13}$ | . | . | . | . | . | . | . |
| **$R_3$** | $b_{14}$ | . | . | . | . | . | . | . |
| **$R_4$** | $b_{15}$ | . | . | . | . | . | . | . |
| **$R_5$** | $b_{16}$ | . | . | . | . | . | . | . |
| **$R_6$** | $b_{17}$ | . | . | . | . | . | . | . |
| **$R_7$** | $b_{18}$ | . | . | . | . | . | . | $b_{88}$ |

Figure 8. *ReC* array before *CMUL* operation.

| CMUL by | $a_{11}$ | $a_{12}$ | $a_{13}$ | $a_{14}$ | $a_{15}$ | $a_{16}$ | $a_{17}$ | $a_{18}$ |
|---|---|---|---|---|---|---|---|---|
| **Columns\|Rows** | **$C_0$** | **$C_1$** | **$C_2$** | **$C_3$** | **$C_4$** | **$C_5$** | **$C_6$** | **$C_7$** |
| **$R_0$** | $a_{11}$**x** $b_{11}$ | . | . | . | . | . | . | . |
| **$R_1$** | $a_{11}$**x** $b_{12}$ | . | . | . | . | . | . | . |
| **$R_2$** | $a_{11}$**x** $b_{13}$ | . | . | . | . | . | . | . |
| **$R_3$** | $a_{11}$**x** $b_{14}$ | . | . | . | . | . | . | . |
| **$R_4$** | $a_{11}$**x** $b_{15}$ | . | . | . | . | . | . | . |
| **$R_5$** | $a_{11}$**x** $b_{16}$ | . | . | . | . | . | . | . |
| **$R_6$** | $a_{11}$**x** $b_{17}$ | . | . | . | . | . | . | . |
| **$R_7$** | $a_{11}$**x** $b_{18}$ | . | . | . | . | . | . | $a_{18}$**x** $b_{88}$ |

Figure 9. *ReC* array after *CMUL* operation.

| Columns\|Rows | $C_0$ | $C_1$ | … |
|---|---|---|---|
| $R_0$ | $a_{11}$x $b_{11}$ | $a_{11}$x $b_{11}$+ $a_{12}$x $b_{21}$ | … |
| $R_1$ | $a_{11}$x $b_{12}$ | . | |
| $R_2$ | $a_{11}$x $b_{13}$ | . | |
| $R_3$ | $a_{11}$x $b_{14}$ | . | |
| $R_4$ | $a_{11}$x $b_{15}$ | . | |
| $R_5$ | $a_{11}$x $b_{16}$ | . | |
| $R_6$ | $a_{11}$x $b_{17}$ | . | |
| $R_7$ | $a_{11}$x $b_{18}$ | . | |

Figure 10. *ReC* array after one accumulation.

After finishing the multiplication of the first row of *A* with the matrix *B*, The results output from the *ReCs* need to be accumulated to produce the first element of matrix *C*. The required *ReC* operation is *CMULOADD* defined as *Out(t + 1) = Out(t) + Out [From Left Cell]*. Finally, the contents of the *ReC* array would be as shown in Figure 11. Note that the general form of this operation is *Out(t +1) = Out(t) + (A x C)*, where *A* is the output from the left cell, and *C* is a constant stored in the context word; here *C* is equal to *1*. Indeed, the contents of column 7 of the *ReC* array are stored back to the frame memory and then to the main memory. The same steps are repeated with the same context word but with different constant field containing the data from matrix *A* until obtaining the resultant matrix *C*. The code and its discussion are in Table 5.

### 4.4.2. Second Mapping

Using the same standard matrix multiplication algorithm, in this section, we introduce a new parallelization scenario, taking the advantages of other supported topologies in the *MorphoSys*. The mapping uses the upper left quadrant along with the bottom right quadrant of the *ReC* array. The matrices are considered to be of size *4x4*. Accordingly, the matrices multiplication finishes with 2 steps. Matrix *B* is broadcasted to the *ReC a*rray, while matrix *A* is stored in the frame buffer. Figure 11 shows the contents of the *ReC* array before any operation. Figure 12 shows the contents of the *ReC* array after the *CMUL* operation. Figure 13 shows the contents of the *ReC* array after one accumulation. The algorithm is to stop after one repetition of the same procedure over the remaining elements in *A* and *B*. The code is discussed in Table 6.

| **Columns\| Rows** | **$C_0$** | **$C_1$** | **$C_2$** | **$C_3$** | **$C_4$** | **$C_5$** | **$C_6$** | **$C_7$** |
|---|---|---|---|---|---|---|---|---|
| **$R_0$** | $b_{11}$ | $b_{21}$ | $b_{31}$ | $b_{41}$ | . | . | . | . |
| **$R_1$** | $b_{12}$ | $b_{22}$ | $b_{32}$ | $b_{42}$ | . | . | . | . |
| **$R_2$** | $b_{13}$ | $b_{23}$ | $b_{33}$ | $b_{43}$ | . | . | . | . |
| **$R_3$** | $b_{14}$ | $b_{24}$ | $b_{34}$ | $b_{44}$ | . | . | . | . |
| **$R_4$** | . | . | . | . | $b_{11}$ | $b_{21}$ | $b_{31}$ | $b_{41}$ |
| **$R_5$** | . | . | . | . | $b_{12}$ | $b_{22}$ | $b_{32}$ | $b_{42}$ |
| **$R_6$** | . | . | . | . | $b_{13}$ | $b_{23}$ | $b_{33}$ | $b_{43}$ |
| **$R_7$** | . | . | . | . | $b_{14}$ | $b_{24}$ | $b_{34}$ | $b_{44}$ |

Figure 11. *ReC* array Before *CMUL* Operation.

| **Columns\| Rows** | **$C_0$** | **$C_1$** | **$C_2$** | **$C_3$** | **$C_4$** | **$C_5$** | **$C_6$** | **$C_7$** |
|---|---|---|---|---|---|---|---|---|
| **$R_0$** | $a_{11}$x $b_{11}$ | … | … | … | . | . | . | . |
| **$R_1$** | $a_{11}$x $b_{12}$ | … | … | … | . | . | . | . |
| **$R_2$** | $a_{11}$x $b_{13}$ | … | … | … | . | . | . | . |
| **$R_3$** | $a_{11}$x $b_{14}$ | … | … | … | . | . | . | . |
| **$R_4$** | . | . | . | . | $a_{11}$x $b_{14}$ | … | … | … |
| **$R_5$** | . | . | . | . | $a_{12}$x $b_{11}$ | … | … | … |
| **$R_6$** | . | . | . | . | $a_{12}$x $b_{12}$ | … | … | … |
| **$R_7$** | . | . | . | . | $a_{12}$x $b_{13}$ | … | … | … |

Figure 12. *ReC* array after one *CMUL* Operation.

| Columns\| Rows | $C_0$ | $C_1$ | $C_2$ | $C_3$ | $C_4$ | $C_5$ | $C_6$ | $C_7$ |
|---|---|---|---|---|---|---|---|---|
| **$R_0$** | $a_{11}$x $b_{11}$ | $a_{11}$x $b_{11}$+ $a_{12}$x $b_{21}$ | … | ... | . | . | . | . |
| **$R_1$** | $a_{11}$x $b_{12}$ | … | … | … | . | . | . | . |
| **$R_2$** | $a_{11}$x $b_{13}$ | … | … | … | . | . | . | . |
| **$R_3$** | $a_{11}$x $b_{14}$ | … | … | … | . | . | . | . |
| **$R_4$** | . | . | . | . | $a_{12}$x $b_{14}$ | $a_{12}$x $b_{11}$+ $a_{12}$x $b_{21}$ | … | … |
| **$R_5$** | . | . | . | . | $a_{12}$x $b_{11}$ | … | … | … |
| **$R_6$** | . | . | . | . | $a_{12}$x $b_{12}$ | … | … | … |
| **$R_7$** | . | . | . | . | $a_{12}$x $b_{13}$ | … | … | … |

Figure 13. *ReC* array after the accumulation.

**Table 5. *TinyRISC* code for a parallel *8x8* matrix multiplication algorithm.**

| | | | |
|---|---|---|---|
| 0: | ldui | r1, 0x1; | R1 ← $10000_{hex}$. This is where matrix B is stored. |
| 1: | ldfb | r1, 0, 0, 16 ; | FB ← 16 x 32 bits at set 0, bank A, address 0. |
| 2: | add | r0, r0, r0; | No-operation. |
| . | . | . | |
| 33: | ldui | r3, 0x3; | R3 ← $30000_{hex}$. This is where the first set of context words is stored in main memory. |
| 34: | ldctxt | r3, 0, 0, 0, 8; | Load 8-context words from main memory starting at the address stored in register 3 into plane 0, block 0 and starting at word 0. for a single row calculation from matrix A. |
| 35: | add | r0, r0, r0; | NOP |
| . | . | . | |
| 38: | sbcb | 0, 0, 0, 0, 0, 0, 0x0; | Single bank column broadcast causing all the cells in the RC array column 0 to perform their operations specified by the context word in context memory starting with data from frame buffer, set 0, bank A [which contain the matrix B elements], address offset 0. |
| 39: | sbcb | 0, 1, 0, 0, 0, 0, 0x40; | It sends data from address $40_{hex}$ in FB.<br>Column 1 in the RC-Array. |
| 40: | sbcb | 0, 2, 0, 0, 0, 0, 0x80; | It sends data from address $80_{hex}$ in FB.<br>Column 2 in the RC-Array. |
| 41: | sbcb | 0, 3, 0, 0, 0, 0, 0xC0; | It sends data from address $C0_{hex}$ in FB |
| 42: | sbcb | 0, 4, 0, 0, 0, 0, 0x100; | It sends data from address $100_{hex}$ in FB |
| 43: | sbcb | 0, 5, 0, 0, 0, 0, 0x140; | It sends data from address $140_{hex}$ in FB |
| 44: | sbcb | 0, 6, 0, 0, 0, 0, 0x180; | It sends data from address $180_{hex}$ in FB |

**Table 5. Continued.**

| | | | |
|---|---|---|---|
| 45: | sbcb | 0, 7, 0, 0, 0, 0, 0x1C0; | It sends data from address $1C0_{hex}$ in FB |
| 46: | ldui | r3, 0x3; | R3 ← $60000_{hex}$. This is where the context word for CMULOADD is stored in main memory. |
| 47: | ldctxt | r3, 0, 0, 8, 1; | Load 1-context word from main memory starting at the address stored in register 3 into plane 8, block 0 and starting at word 0. |
| 48: | add | r0, r0, r0; | NOP |
| . | . | . | |
| 51: | ldui | r4, 0x0; | R4 ← $00000_{hex}$. |
| 52: | sbcb | 0,1, 0, 0, 0, 0, 0x40; | Single bank column broadcast causing all the cells in the RC array column 0 to perform their operations specified by the context word in context memory with data from the left cell output. |
| 53: | sbcb | 0,2, 0, 0, 0, 0, 0x80; | |
| 54: | sbcb | 0,3, 0, 0, 0, 0, 0xC0; | |
| 55: | sbcb | 0,4, 0, 0, 0, 0, 0x100; | |
| 56: | sbcb | 0,5, 0, 0, 0, 0, 0x140; | |
| 57: | sbcb | 0,6, 0, 0, 0, 0, 0x180; | |
| 58: | sbcb | 0,7, 0, 0, 0, 0, 0x1C0; | |
| 59: | wfbi | 7, 0, 0, 1, 0x0; | Write data back to the frame buffer from the output registers of column 7 into set 1, address 0. |
| 248: | Repeat the above code 7 times, with the appropriate memory shifts inside the instructions this repetition would take additional 189 cycles. | | |

**Table 6. *TinyRISC* code for a parallel *4x4* matrix multiplication algorithm.**

| | | | |
|---|---|---|---|
| 0: | ldui | r1, 0x1; | R1 ← $10000_{hex}$. This is where matrix B is stored. |
| 1: | ldfb | r1, 0, 0, 16 ; | FB ← 4 x 32 bits at set 0, bank A, address 0. |
| 2: | add | r0, r0, r0; | No-operation. |
| . | . | . | |
| 9: | ldui | r3, 0x3; | R3 ← $30000_{hex}$. This is where the first set of context words is stored in main memory. |
| 10: | ldctxt | r3, 0, 0, 0, 8; | Load 8-context words from main memory starting at the address stored in register 3 into plane 0, block 0 and starting at word 0. for a single row calculation from matrix A. |
| 11: | add | r0, r0, r0; | NOP |
| . | . | . | |
| 14: | sbcb | 0, 0, 0, 0, 0, 0, 0x0; | Single bank column broadcast causing all the cells in the RC array column 0 to perform their operations specified by the context word in context memory starting with data from frame buffer, set 0, bank A [which contain the matrix B elements], address offset 0. |

**Table 6. Continued.**

| | | | |
|---|---|---|---|
| 15: | sbcb | 0, 1, 0, 0, 0, 0, 0x20; | It sends data from address $20_{hex}$ in FB.<br>Column 1 in the RC-Array. |
| 16: | sbcb | 0, 2, 0, 0, 0, 0, 0x40; | It sends data from address $40_{hex}$ in FB.<br>Column 2 in the RC-Array. |
| 17: | sbcb | 0, 3, 0, 0, 0, 0, 0x60; | It sends data from address $60_{hex}$ in FB |
| 18: | sbcb | 0, 4, 0, 0, 0, 0, 0x0; | It sends data from address $0_{hex}$ in FB |
| 19: | sbcb | 0, 5, 0, 0, 0, 0, 0x20; | It sends data from address $20_{hex}$ in FB |
| 20: | sbcb | 0, 6, 0, 0, 0, 0, 0x40; | It sends data from address $40_{hex}$ in FB |
| 21: | sbcb | 0, 7, 0, 0, 0, 0, 0x60; | It sends data from address $60_{hex}$ in FB |
| 22: | ldui | r3, 0x3; | R3 $\leftarrow$ $60000_{hex}$. This is where the context word for CMULOADD is stored in main memory. |
| 23: | ldctxt | r3, 0, 0, 8, 1; | Load 1-context word from main memory starting at the address stored in register 3 into plane 8, block 0 and starting at word 0. |
| 24: | add | r0, r0, r0; | NOP |
| . | . | . | |
| 27: | ldui | r4, 0x0; | R4 $\leftarrow$ $00000_{hex}$. |
| 28: | sbcb | 0,1, 0, 0, 0, 0, 0x0; | Single bank column broadcast causing all the cells in the RC array column 1 to perform their operations specified by the context word in context memory with data from the left cell output. |
| 29: | sbcb | 0,2, 0, 0, 0, 0, 0x00; | |
| 30: | sbcb | 0,3, 0, 0, 0, 0, 0x00; | |
| 31: | sbcb | 0,4, 0, 0, 0, 0, 0x00; | |
| 32: | sbcb | 0,5, 0, 0, 0, 0, 0x00; | |
| 33: | sbcb | 0,6, 0, 0, 0, 0, 0x00; | |
| 34: | sbcb | 0,7, 0, 0, 0, 0, 0x00; | |
| 35: | wfbi | 7, 0, 0, 1, 0x0; | Write data back to the frame buffer from the output registers of column 7 into set 1, address 0. |
| 70: | Repeat the above code once, with the appropriate memory shifts inside the instructions this repetition would take additional 35 cycles. | | |

## 4.5. Performance Evaluation and Analysis

The performance evaluation is based on the speed of execution of the developed parallel algorithms. The speed of execution is calculated based on the number of cycles taken by a program under *MorphoSys* to execute. The *MorphoSys* system is considered to be operational at a frequency of 100 *MHz*.

In Table 7 the results are shown including the number of elements, number of cycles, Execution Time in μsec, the number of elements processed by cycle, and number of cycles taken to produce an element from the desired output.

**Table 7. Summary of Findings.**

| **Algorithm** | **No. of Elements** | **No. of Cycles** | **Execution Time in µsec** | **Elements per Cycle** | **Cycles per** |
|---|---|---|---|---|---|
| Translation | **8** | 21 | 0.21 | 0.38 | 2.625 |
| Translation | **64** | 96 | 0.96 | 0.667 | 1.5 |
| Composite Translation | **8** | 44 | 0.44 | 0.18 | 5.5 |
| Composite Translation | **64** | 224 | 2.24 | 0.285 | 3.5 |
| Scaling | **8** | 14 | 0.14 | 0.57 | 1.75 |
| Scaling | **64** | 55 | 0.55 | 1.16 | 0.859 |
| Composite Scaling | **8** | 23 | 0.23 | 0.35 | 2.875 |
| Composite Scaling | **64** | 66 | 0.66 | 0.97 | 1.03 |
| Composition of Translation and Scaling Operations | **8** | 21 | 0.21 | 0.380 | 2.625 |
| Composition of Translation and Scaling Operations | **64** | 151 | 1.51 | 0.42 | 2.36 |
| Transformations using Matrix Multiplication Algorithm I | **64** | 248 | 2.48 | 0.258 | 3.875 |
| Transformations using Matrix Multiplication Algorithm II | **16** | 70 | 0.7 | 0.228 | 4.375 |

With a cycle count of 21, the number of elements per cycle for the 8-element translation algorithm is 0.38. The number of elements per cycle measure for a 64-element translation is 0.667, with its 96 cycles. Accordingly, the maximum utilization of the *ReC* array has lead to a higher throughput. The same conclusion is reached throughout all the algorithms mapped; the higher the utilization of the *ReC* array, then the higher the throughput.

The composite routines (translation and scaling operations) achieves a speedup over performing the same composite operations by running a single translation (or scaling operation) twice. Table 8 shows the speedup of the composite routines over the standard basic routines. The speedup is calculated as the ratio in number of cycles.

**Table 8. Comparisons among different translation and scaling operation mappings.**

| Algorithm | No. of Elements | No. of Cycles | Speedup | Element Per Cycles | Cycles Per Element |
|---|---|---|---|---|---|
| **Translation** | | | | | |
| Translating twice using two calls of the translation routine | 8 | 42 | **1.35** | 0.19 | 5.2 |
| Translating twice using the composite translation routine | 8 | 31 | | 0.258 | 3.075 |
| Translating twice using two calls of the translation routine | 64 | 192 | **1.28** | 0.33 | 3 |
| Translating twice using the composite translation routine | 64 | 150 | | 0.42 | 2.34 |
| **Scaling** | | | | | |
| Scaling twice using two calls of the scaling routine | 8 | 28 | **1.2** | 0.285 | 3.5 |
| Scaling twice using the composite scaling routine | 8 | 23 | | 0.35 | 2.875 |
| Scaling twice using two calls of the scaling routine | 64 | 110 | **1.67** | 0.58 | 1.78 |
| Scaling twice using the composite scaling routine | 64 | 66 | | 0.97 | 1.03 |
| **Translation and Scaling** | | | | | |
| Executing the translation routine then scaling routine | 8 | 35 | **1.67** | 0.228 | 4.38 |
| Translation and scaling in a single routine | 8 | 21 | | 0.25 | 3.88 |
| Executing the translation routine then scaling routine | 64 | 151 | **1.57** | 0.42 | 2.35 |
| Translation and scaling in a single routine | 64 | 96 | | 0.42 | 0.34 |

The developed parallel matrix multiplication algorithms allows for an increased flexibility in performing geometrical transformations. The matrix representation is a general representation to implement all basic and composite geometrical transformations; this is one of the key advantages. From performance point of view, the parallel matrix multiplication algorithms might not be apparently faster than those designed for specific transformations. The matrix multiplication-based transformation algorithms become faster when the number of calls of basic transformations becomes large. For example, the matrix-based representation, and after calculating the composite translation constant, is more efficient than running the basic translation routine a large number of times. The matrix form will allow applying the matrix multiplication routine once with a single accumulated translation value. The same scenario is true in the case of scaling, rotation, and shearing.

# 5. Cyclic Redundancy Checkers under *MorphoSys*

Redundant encoding is a method of error detection that spreads the information across more bits than the original data. The more redundant bits used, the greater the chance to detect errors. *CRCCs* are check for differences between transmitted data and the original data. *CRCCs* are effective for two reasons: Firstly, they provide excellent protection against common errors, such as burst errors where consecutive bits in a data stream are corrupted during transmission. Secondly, systems that use *CRCCs* are easy to implement [72, 75, 76]. When a *CRCC* is used to verify a frame of data, the frame is treated as one very large binary number, which is then divided by a generator number. This division produces a reminder, which is transmitted along with the data. At the receiving end, the data is divided by the same generator number and the remainder is compared with the one sent at the end of the data frame. If the two remainders are different, then an error occurred during data transmission. Types of errors that a *CRCC* detects depend on the generator polynomial. Table 9 shows the most common generator polynomials.

**Table 9. Common generator polynomials.**

| Generator | Polynomial |
|---|---|
| SDLC (CCITT) | $X^{16} + X^{12} + X^5 + X^0$ |
| SDLC Reverse | $X^{16} + X^{11} + X^4 + X^0$ |
| CRC-16 | $X^{16} + X^{15} + X^2 + X^0$ |
| CRC-16 Reverse | $X^{16} + X^{14} + X^1 + X^0$ |
| CRC-12 | $X^{12} + X^{11} + X^3 + X^2 + X^1 + X^0$ |
| Ethernet | $X^{32} + X^{26} + X^{23} + X^{22} + X^{16} + X^{12} + X^{11} + X^{10} + X^8 + X^7 + X^5 + X^4 + X^2 + X^1 + X^0$ |

## 5.1. CRC Sequential Implementation

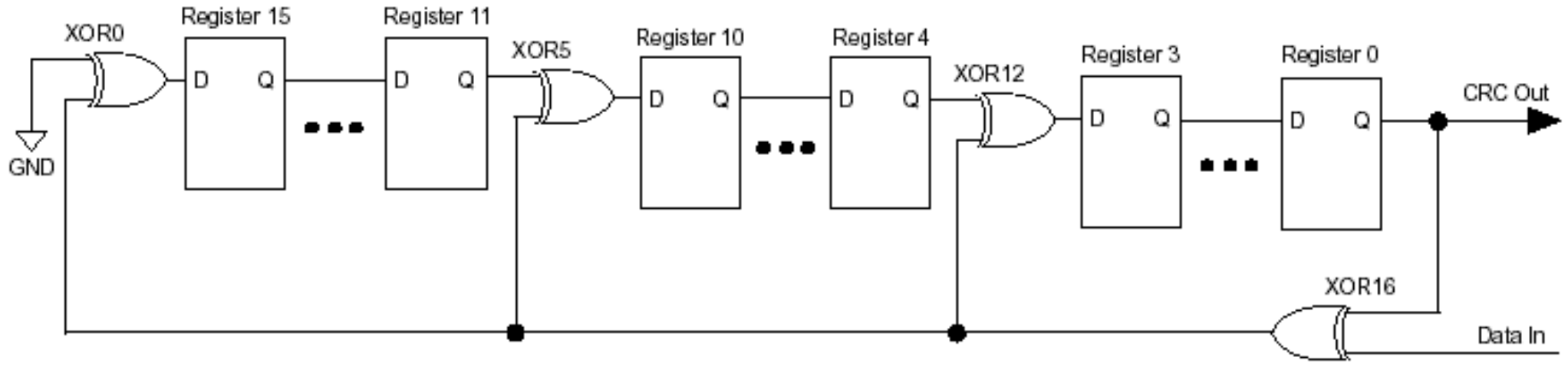

Figure 14. *LFSR* implementation of the *CCITT CRC-16*.

CRC implementation is usually done with linear-feedback shift registers (LFSRs). Figures 14 and 15 show the *CCITT CRC-16* and *CRC-16* generators with their serial implementation using *LFSRs*. This serial method works well when the data is available in bit-stream form.

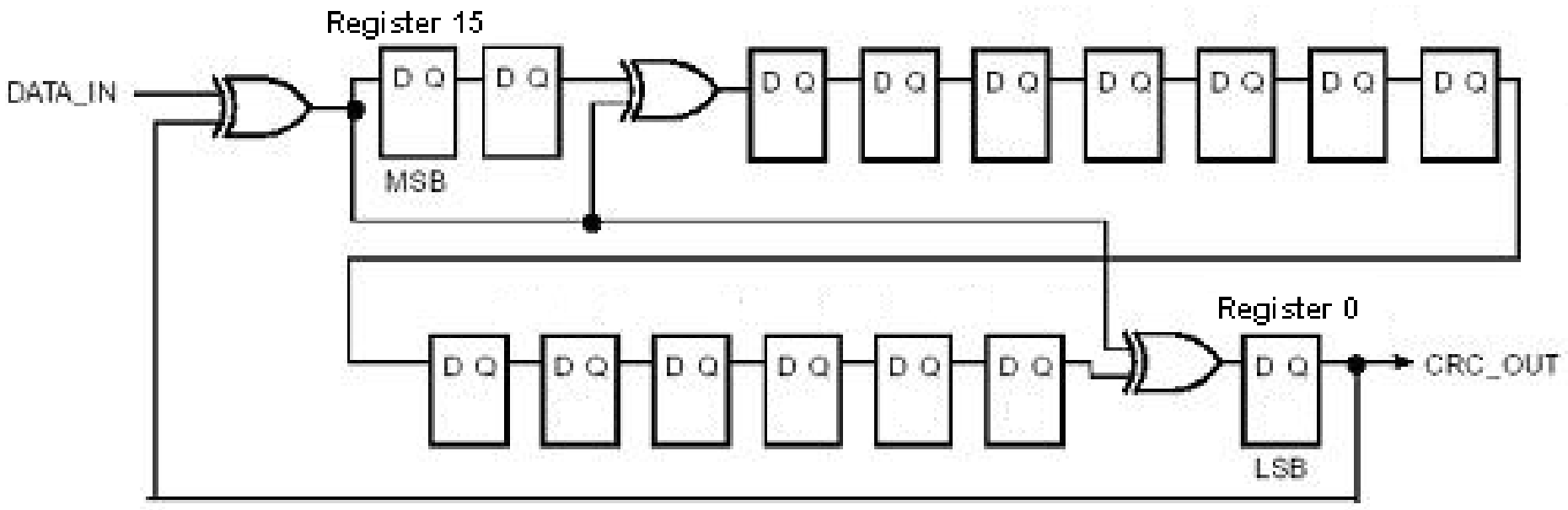

Figure 15. *LFSR* implementation of the *CRC-16*.

## 5.2. CRC Parallel Implementation

With the currently available high-speed digital signal processing (*DSP*) systems, the processing of data is done in a byte, word, double word, or larger widths rather than serially. Even with serial telecommunication systems, data is buffered in chips responsible for synchronizations and framing. For parallel implementation, the data is available in 8-bit frames with manageable speed [72]. A one channel parallel *CRC* algorithm with *LFSR* approach is done by considering the state of the circuit on 8-shifts basis [73]. Tables 10 and 11, show two different implementations of the *CCITT CRC-16* and *CRC-16*. The term $Register_i$ represents the *LFSR* internal register number "*i*", while *XORj* represents the output of the *XOR* gate number "*j*", and *XOR* indicates the *exclusive-OR* operation. With the emergence of the highly scalable reconfigurable circuits, more implementation capabilities are present. Along with the byte-wise or word-wise *CRC* implementation, it is possible to implement parallel channels each with byte-wise *CRC* implementation.

**Table 10. The states of registers after 8-shifts for the *CCITT CRC-16* Algorithm.**

| New Values After 8-shifts of the registers and the output of the XOR-gates | |
|---|---|
| $XOR_i$ = | $Register_i \oplus DataIn_i$ i = 0, 1, …, 7 |
| | |
| $Register_0$ = | $Register_8 \oplus XOR_4 \oplus XOR_0$ |
| $Register_1$ = | $Register_9 \oplus XOR_5 \oplus XOR_1$ |
| $Register_2$ = | $Register_{10} \oplus XOR_6 \oplus XOR_2$ |
| $Register_3$ = | $Register_{11} \oplus XOR_0 \oplus XOR_7 \oplus XOR_3$ |
| $Register_4$ = | $Register_{12} \oplus XOR_1$ |
| $Register_5$ = | $Register_{13} \oplus XOR_2$ |
| $Register_6$ = | $Register_{14} \oplus XOR_3$ |
| $Register_7$ = | $Register_{15} \oplus XOR_4 \oplus XOR_0$ |
| $Register_8$ = | $XOR_0 \oplus XOR_5 \oplus XOR_1$ |
| $Register_9$ = | $XOR_1 \oplus XOR_6 \oplus XOR_2$ |
| $Register_{10}$ = | $XOR_2 \oplus XOR_7 \oplus XOR_3$ |
| $Register_{11}$ = | $XOR_3$ |
| $Register_{12}$ = | $XOR_4 \oplus XOR_0$ |
| $Register_{13}$ = | $XOR_5 \oplus XOR_1$ |
| $Register_{14}$ = | $XOR_6 \oplus XOR_2$ |
| $Register_{15}$ = | $XOR_7 \oplus XOR_3$ |

**Table 11. The states of the registers after 8-shifts for the *CRC-16* Algorithm.**

| New Values After 8-shifts of the registers and the output of the XOR-gates | |
|---|---|
| $XOR_i$ = | $Register_i \oplus DataIn_i$ $i = 0, 1, \ldots, 7$ |
| | |
| X | $XOR_0 \oplus XOR_1 \oplus \ldots XOR_7$ |
| | |
| $Register_0$ = | $Register_8 \oplus X$ |
| $Register_1$ = | $Register_9$ |
| $Register_2$ = | $Register_{10}$ |
| $Register_3$ = | $Register_{11}$ |
| $Register_4$ = | $Register_{12}$ |
| $Register_5$ = | $Register_{13}$ |
| $Register_6$ = | $Register_{14} \oplus XOR_0$ |
| $Register_7$ = | $Register_{15} \oplus XOR_1 \oplus XOR_0$ |
| $Register_8$ = | $XOR_3 \oplus XOR_2$ |
| $Register_9$ = | $XOR_4 \oplus XOR_3$ |
| $Register_{10}$ = | $XOR_5 \oplus XOR_4$ |
| $Register_{11}$ = | $XOR_6 \oplus XOR_5$ |
| $Register_{12}$ = | $XOR_7 \oplus XOR_6$ |
| $Register_{13}$ = | $XOR_8 \oplus XOR_7$ |
| $Register_{14}$ = | $XOR_8 \oplus X$ |
| $Register_{15}$ = | X |

## 5.3. Mapping CRC Algorithms

From the underlying architecture point of view, the mapping of any algorithm onto the proposed reconfigurable system requires in-depth knowledge of all the available interconnection topologies. Moreover, the designer should take into consideration the possibility of dynamically changing the shape of the interconnection. From the algorithmic point of view, the design of a parallel version of the addressed algorithms requires the best use of recourses with the least possible redundancy in computations.

### 5.3.1. The CCITT CRC-16 Algorithm Mapping

The mapping of the parallel *CCITT CRC-16* algorithm will make use of the redundant computations utilized in several steps of the algorithm. Firstly, the values of $XOR_i$ for all values of *i* from 0 to 7 are calculated. In Table 10 the computations that are used more than once are shown. Particularly, the redundant values are $(XOR_{i+4} \oplus XOR_i)$ for *i* from *0* to *3*. Thus, the second computation step involves registers 4, 5, 6, 11, 12, 13, 14, and 15 depending on results found in the first step. In the final step the computations for registers 0, 1, 2, 3, 7, 8, 9 and 10 are carried out depending on the results calculated in the first and second step.

The algorithm mapping will be explained by introducing the three needed sets of code. The first set is the interconnection context words. The context word used in this algorithm is that for *XOR* with column broadcast, where each cell *XORs* two inputs from frame buffers *A* and *B*. This context word is stored at address $30000_{hex}$.

The second set of code is the input data and the initial data in the circuits registers. These two sets of data are stored in main memory address $10000_{hex}$ and $20000_{hex}$.

The third set is that of the *TinyRISC* code, which the main code is. This code and its discussion are shown in Table 12. Main steps of the addressed algorithm are shown in Figures 16 and 17. The final contents of frame buffer *A* is shown in Figure 18.

**Table 12. The *TinyRISC* code for the *CRC CCITT-16* Algorithm.**

| | | | |
|---|---|---|---|
| 0: | ldui | r1, 0x1; | R1 ← $10000_{hex}$. This is where DataIn Stored |
| 1: | ldfb | r1, 0, 0, 2 ; | FB ← 2 x 32 bits at set 0, bank A, address 0. |
| 2: | add | r0, r0, r0; | No-operation. |
| . | . | . | |
| 6: | ldui | r2, 0x1; | R2 ← $20000_{hex}$. This is where Circuit Registers Stored. |
| 7: | ldfb | r2, 1, 0, 2 ; | FB ← 2 x 32 bits at set 0, bank B, address 0. |
| 8: | add | r0, r0, r0; | No-operation. |
| . | . | . | |
| 12: | ldui | r3, 0x3; | R3 ← $30000_{hex}$. This is where the context word is stored in main memory. |
| 7: | ldctxt | r3, 0, 0, 0, 1; | Load 1 context word from main memory starting at the address stored in register 3 into plane 0, block 0 and starting at word 0. |
| 8: | add | r0, r0, r0; | NOP |
| . | . | . | |
| 11: | ldui | r4, 0x0; | R4 ← 00000hex. |
| 12: | dbcdc | r4, 0, 0, 0, 0, 0, 0; | Double bank column broadcast. It sends data from both banks address 0 in the FB and broadcasts the context words column-wise. It triggers the RC array to start execution of column 0 by the context word of address 0 in the column block of context memory operating on data in set 0. Bank A starting at 0. Bank B starting at (0 + 0). |
| 13: | wfbi | 0, 0, 0, 0, 0x0; | Write data back to the FB A from the output registers of column 0 into set 0, address 0. Results for $XOR_i$. The value of $Register_{11}$. |
| 14: | wfbi | 0, 0, 1, 0, $10_{hex}$; | Write data back to FB B from the output registers of column 0 into set 0, address $10_{hex}$. |
| 15: | dbcdc | r4, $14_{hex}$, 0, 0, 0, 1, 0; | Double bank column broadcast. Bank A starting at 0x0. Bank B starting at (0x0 + 14), i.e. shifted 4 words from the starting point of $XOR_0$. The calculated items in this operation are the repeatedly used operations shaded in Table 1. |
| 16: | wfbi | 0, 0, 0, 0, $30_{hex}$; | Write data back to the FB from the output registers of column 0 into set 0, address $1D_{hex}$. Values for Circuits Registers [12..15] are now available in FB A starting from address $30_{hex}$. |
| 17: | dbcdc | r4, $8_{hex}$, 0, 0, 0, 0, $30_{hex}$; | Double bank column 0 broadcast. Bank A starting at $30_{hex}$. Bank B starting at (0x0 + $8_{hex}$). Thus, Registers [0..2] values are now available. |
| 18: | dbcdc | r4, $C_{hex}$, 0, 1, 0, 0, $1_{hex}$; | Double bank column 1 broadcast. Bank A starting at $1_{hex}$. Bank B starting at (0x0 + $C_{hex}$). Thus, Registers [4..6] values are now available. |
| 19: | dbcdc | r4, $10_{hex}$, 0, 2, 0, 0, $31_{hex}$; | Double bank column 2 broadcast Bank A starting at $30_{hex}$. Bank B starting at (0x0 + $10_{hex}$). Thus, Registers [8..10] values are now available. |

**Table 12. Continued.**

| | | | |
|---|---|---|---|
| 20: | wfbi | 0, 0, 0, 0, $40_{hex}$; | Write data back to the frame buffer A from the output registers of column 1 into set 0, address $40_{hex}$. Values for Circuits Registers [0..2]. |
| 21: | wfbi | 1, 0, 0, 0, $50_{hex}$; | Write data back to the FB A from the output registers of column 2 into set 0, address $50_{hex}$. Values for Circuits Registers [4..6]. |
| 22: | wfbi | 2, 0, 0, 0, $55_{hex}$; | Write data back to the frame buffer A from the output registers of column 0 into set 0, address $55_{hex}$. Values for Circuits Registers [8..10]. |
| 23: | dbcdc | r4, $10_{hex}$, 0, 0, 0, 0, $43_{hex}$; | Double bank column 0 broadcast. Bank A starting at $43_{hex}$. Bank B starting at (0x0 + $10_{hex}$). The output calculated here is the xor operation of $XOR_0$, $XOR_7$, & $XOR_3$ the value is in cell 0 c 0. |
| 22: | wfbi | 0, 0, 0, 0, $60_{hex}$; | Write data back to the FB A from the output registers of column 0 into set 0, address $60_{hex}$. |
| 23: | dbcdc | r4, $C_{hex}$, 0, 0, 0, 0, $60_{hex}$; | Double bank column 2 broadcast. Bank A starting at $60_{hex}$. Bank B starting at (0x0 + $C_{hex}$). Thus, $Register_3$ value is now calculated. |
| 24: | wfbi | 0, 0, 0, 0, $60_{hex}$; | Write data back to the FB A from the output registers of column 0 into set 0, address $60_{hex}$. |
| 25: | dbcdc | r4, $10_{hex}$, 0, 0, 0, 0, $30_{hex}$; | Double bank column 2 broadcast Bank A starting at $30_{hex}$. Bank B starting at (0x0 + $10_{hex}$). Thus, $Register_7$ value is now calculated. |
| 26: | wfbi | 0, 0, 0, 0, $65_{hex}$; | Write data back to the FB A from the output registers of column 0 into set 0, address $65_{hex}$. |

| **Rows \|Columns** | **$C_0$** | **$C_1$** | **$C_2$** | **$C_3$** | **$C_4$** | **$C_5$** | **$C_6$** | **$C_7$** |
|---|---|---|---|---|---|---|---|---|
| **$R_0$** | $Register_0 \oplus DataIn_0$ | . | . | . | . | . | . | . |
| **$R_1$** | $Register_1 \oplus DataIn_1$ | . | . | . | . | . | . | . |
| **$R_2$** | $Register_2 \oplus DataIn_2$ | . | . | . | . | . | . | . |
| **$R_3$** | $Register_3 \oplus DataIn_3$ | . | . | . | . | . | . | . |
| **$R_4$** | $Register_4 \oplus DataIn_4$ | . | . | . | . | . | . | . |
| **$R_5$** | $Register_5 \oplus DataIn_5$ | . | . | . | . | . | . | . |
| **$R_6$** | $Register_6 \oplus DataIn_6$ | . | . | . | . | . | . | . |
| **$R_7$** | $Register_7 \oplus DataIn_7$ | . | . | . | . | . | . | . |

Figure 16. *ReC* array contents after calculating for $XOR_i$.

| **Rows \|Columns** | **$C_0$** | **$C_1$** | **$C_2$** | **$C_3$** | **$C_4$** | **$C_5$** | **$C_6$** | **$C_7$** |
|---|---|---|---|---|---|---|---|---|
| **$R_0$** | $XOR_4 \oplus XOR_0$ | . | . | . | . | . | . | . |
| **$R_1$** | $XOR_5 \oplus XOR_1$ | . | . | . | . | . | . | . |
| **$R_2$** | $XOR_6 \oplus XOR_2$ | . | . | . | . | . | . | . |
| **$R_3$** | $XOR_7 \oplus XOR_3$ | . | . | . | . | . | . | . |
| **$R_4$** | . | . | . | . | . | . | . | . |
| **$R_5$** | . | . | . | . | . | . | . | . |
| **$R_6$** | . | . | . | . | . | . | . | . |
| **$R_7$** | . | . | . | . | . | . | . | . |

Figure 17. *ReC* array contents for the second computation step for *CRC-CCITT-16*.

| Address In HEX | Frame Buffer A | Address In HEX | Frame Buffer A |
|---|---|---|---|
| 0 | $DataIn_0$ | 40 | **$Register_0$** |
| 1 | $DataIn_1$ | 41 | **$Register_1$** |
| 2 | $DataIn_2$ | 42 | **$Register_2$** |
| 3 | $DataIn_3$ | . | . |
| 4 | $DataIn_4$ | 50 | **$Register_4$** |
| 5 | $DataIn_5$ | 51 | **$Register_5$** |
| 6 | $DataIn_6$ | 52 | **$Register_6$** |
| 7 | $DataIn_7$ | . | . |
| . | . | 55 | **$Register_8$** |
| 10 | $XOR_0$ | 56 | **$Register_9$** |
| 11 | $XOR_1$ | 57 | **$Register_{10}$** |
| 12 | $XOR_2$ | . | |
| 13 | **$XOR_3$ / $Register_{11}$** | 60 | **$Register_3$** |
| 14 | $XOR_4$ | . | |
| 15 | $XOR_5$ | 65 | **$Register_7$** |
| 16 | $XOR_6$ | . | . |
| 17 | $XOR_7$ | . | . |
| . | . | . | . |
| 30 | **$XOR_4 \oplus XOR_0$/ $Register_{12}$** | . | . |
| 31 | **$XOR_5 \oplus XOR_1$/ $Register_{13}$** | . | . |
| 32 | **$XOR_6 \oplus XOR_2$/ $Register_{14}$** | . | . |
| 33 | **$XOR_7 \oplus XOR_3$/ $Register_{15}$** | . | . |

Figure 18. Contents of Frame Buffer *A* after the algorithm terminates after one computation step the new registers values are shown at the specified locations.

### 5.3.2. The CRC-16 Algorithm Mapping

The mapping of this algorithm depends also on eliminating redundant computations, besides, the parallel computation of the required values. This mapping is of three steps. Firstly, the values of $XOR_i$ for all values of $i$ from 0 to 7 are calculated. From Table 11, the computations used more than once are shaded, particularly, the redundant value $X$ $(XOR_0 \oplus \ldots \oplus XOR_7)$. Thus, the second computation step is for $X$. Thirdly, the rest of the values are calculated in parallel.

**Table 13. The *TinyRISC* code for the *CRC-16* Algorithm.**

| | | |
|---|---|---|
| 0: | ldui r1, 0x1; | R1 ← $10000_{hex}$. This is where DataIn Stored |
| 1: | ldfb r1, 0, 0, 2 ; | FB ← 2 x 32 bits at set 0, bank A, address 0. |
| 2: | add r0, r0, r0; | No-operation. |
| . | . . | |
| 6: | ldui r2, 0x1; | R2 ← $20000_{hex}$. This is where Circuit Registers Stored. |
| 7: | ldfb r2, 1, 0, 2 ; | FB ← 2 x 32 bits at set 0, bank B, address 0. |

**Table 13. Continued.**

| | | | |
|---|---|---|---|
| 8: | add | r0, r0, r0; | No-operation. |
| . | . | . | |
| 12: | ldui | r3, 0x3; | R3 ← $30000_{hex}$. This is where the context word is stored in main memory. |
| 7: | ldctxt | r3, 0, 0, 0, 3; | Load 3 context words from main memory starting at the address stored in register 3 into plane 0, block 0 and starting at word 0. |
| 8: | add | r0, r0, r0; | NOP |
| . | . | . | |
| 11: | ldui | r4, 0x0; | R4 ← 00000hex. |
| 12: | dbcdc | r4, 0, 0, 0, 0, 0, 0; | Double bank column broadcast. It sends data from both banks address 0 in the frame buffer and broadcasts the context words column-wise. It triggers the RC array to start execution of column 0 by the context word of address 0 in the column block of context memory operating on data in set 0. Bank A starting at 0x0. Bank B starting at (0x0 + 0). |
| 13: | wfbi | 0, 0, 0, 0, 0x0; | Write data back to the frame buffer A from the output registers of column 0 into set 0, address 0. Results for $XOR_i$. |
| 14: | wfbi | 0, 0, 1, 0, $10_{hex}$; | Write data back to FB B from the output registers of column 0 into set 0, address $10_{hex}$. |
| 15: | sbcb | 0, 0, 0, 1, 0, 0, $0_{hex}$; | Single bank column broadcast causing all the cells in the RC array column 0 to perform their operations specified by the second context word in context memory starting with data from frame buffer, set 0, bank A |
| 16: | sbcb | 0, 1, 0, 1, 0, 0, $1_{hex}$; | It sends data from address $1_{hex}$ in FB Column 1. |
| 17: | sbcb | 0, 2, 0, 1, 0, 0, $2_{hex}$; | It sends data from address $2_{hex}$ in FB Column 2. |
| 18: | sbcb | 0, 3, 0, 1, 0, 0, $3_{hex}$; | It sends data from address $3_{hex}$ in FB. |
| 19: | sbcb | 0, 4, 0, 1, 0, 0, $4_{hex}$; | It sends data from address $4_{hex}$ in FB. |
| 20: | sbcb | 0, 5, 0, 1, 0, 0, $5_{hex}$; | It sends data from address $5_{hex}$ in FB. |
| 21: | sbcb | 0, 6, 0, 1, 0, 0, $6_{hex}$; | It sends data from address $6_{hex}$ in FB, the new value of $Register_{14}$. |
| 22: | sbcb | 0, 7, 0, 1, 0, 0, $7_{hex}$; | It sends data from address $7_{hex}$ in FB, the new value of $Register_{15}$. |
| 23: | wfbi | 6, 0, 0, 0, $30_{hex}$; | Write data back to FB A from the output registers of column 6 into set 0, address $30_{hex}$. Value of $Register_{14}$. |
| 24: | wfbi | 7, 0, 0, 0, $31_{hex}$; | Write data back to the FB A from the output registers of column 7 into set 0, address $31_{hex}$. Value of $Register_{15}$. |

**Table 13. Continued**

| | | | |
|---|---|---|---|
| 25: | dbcdc | r4, $13_{hex}$, 0, 0, 0, 0, $2_{hex}$; | Double bank column broadcast. Bank A starting at $2_{hex}$. Bank B starting at $13_{hex}$. New values for Registers [8..13] are now available. |
| 26: | wfbi | 0, 0, 0, 0, $35_{hex}$; | Write data back to the FB A from the output registers of column 0 into set 0, address $35_{hex}$. |
| 27: | dbcdc | r4, $8_{hex}$, 0, 0, 0, 0, $31_{hex}$; | Double bank column broadcast. It sends data from Bank A starting at $31_{hex}$. Bank B starting at $8_{hex}$. New value for $Register_0$. |
| 28: | wfbi | 0, 0, 0, 0, $45_{hex}$; | Write data back to the FB A from the output registers of column 0 into set 0, address $45_{hex}$. |
| 29: | dbcdc | r4, $14_{hex}$, 0, 0, 0, 0, $0_{hex}$; | Double bank column broadcast. It sends data from Bank A starting at $0_{hex}$. Bank B starting at $14_{hex}$. New values for $Register_6$. |
| 30: | wfbi | 0, 0, 0, 0, $50_{hex}$; | Write data back to the FB A from the output registers of column 0 into set 0, address $50_{hex}$. |
| 31: | dbcdc | r4, $10_{hex}$, 0, 0, 0, 0, $35_{hex}$; | Double bank column broadcast. It sends data from Bank A starting at $2_{hex}$. Bank B starting at $13_{hex}$. New values for $Register_7$. |
| 30: | wfbi | 0, 0, 0, 0, $55_{hex}$; | Write data back to the FB A from the output registers of column 0 into set 0, address $45_{hex}$. |

The algorithm mapping will be explained by introducing the three needed sets of code. The first set is the interconnection context words. The context words used in this algorithm are firstly, that for *XOR* with column broadcast, where each cell *XORs* two inputs from frame buffers *A* and *B*. Secondly, the same cell operation is used also by taking one input from the frame buffer, and the second from the output of the left adjacent cell. The context words are stored at address $30000_{hex}$. The second set of code is the input data and the initial data in the circuits registers. These two sets of data are stored in main memory address $10000_{hex}$ and $20000_{hex}$. The third set is that of the *TinyRISC* code which is the main code, this code and its discussion are shown in Table 13. Main steps of the addressed algorithm are shown in Figures 19 and 20. The final contents of frame buffer *A* is shown in Figure 21.

| **Rows \|Columns** | **$C_0$** | **…** | **$C_7$** |
|---|---|---|---|
| **$R_0$** | $Register_0 \oplus DataIn_0$ | . | . |
| **$R_1$** | $Register_1 \oplus DataIn_1$ | . | . |
| **$R_2$** | $Register_2 \oplus DataIn_2$ | . | . |
| **$R_3$** | $Register_3 \oplus DataIn_3$ | . | . |
| **$R_4$** | $Register_4 \oplus DataIn_4$ | . | . |
| **$R_5$** | $Register_5 \oplus DataIn_5$ | . | . |
| **$R_6$** | $Register_6 \oplus DataIn_6$ | . | . |
| **$R_7$** | $Register_7 \oplus DataIn_7$ | . | . |

Figure 19. *ReC* array contents after calculating for $XOR_i$.

| Rows \|Columns | $C_0$ | $C_1$ | $C_2$ | $C_3$ | $C_4$ | $C_5$ | $C_6$ | $C_7$ |
|---|---|---|---|---|---|---|---|---|
| **$R_0$** | $XOR_0$ | $XOR_0 \oplus XOR_1$ | . | . | . | . | . | $XOR_0 \oplus \ldots \oplus XOR_7$ |
| **$R_1$** | . | . | . | . | . | . | . | . |
| **$R_2$** | . | . | . | . | . | . | . | . |
| **$R_3$** | | | | | | | | |
| **$R_4$** | | | | | | | | |
| **$R_5$** | | | | | | | | |
| **$R_6$** | | | | | | | | |
| **$R_7$** | | | | | | | | |

Figure 20. *ReC* array contents for the second computation step for *CRC-16*.

| Address In HEX | Frame Buffer A | Address In HEX | Frame Buffer A |
|---|---|---|---|
| 0 | $DataIn_0$ | . | . |
| 1 | $DataIn_1$ | 36 | **$Register_8$** |
| 2 | $DataIn_2$ | 37 | **$Register_9$** |
| 3 | $DataIn_3$ | 38 | **$Register_{10}$** |
| 4 | $DataIn_4$ | 39 | **$Register_{11}$** |
| 5 | $DataIn_5$ | 40 | **$Register_{12}$** |

Figure 21. Continued on next page.

| Address In HEX | Frame Buffer A | Address In HEX | Frame Buffer A |
|---|---|---|---|
| 6 | $DataIn_6$ | 41 | **$Register_{13}$** |
| 7 | $DataIn_7$ | . | . |
| . | . | 45 | **$Register_0$** |
| 10 | $XOR_0$ | . | . |
| 11 | $XOR_1$ | 50 | **$Register_6$** |
| 12 | $XOR_2$ | . | . |
| 13 | $XOR_3$ | 55 | **$Register_7$** |
| 14 | $XOR_4$ | . | . |
| 15 | $XOR_5$ | . | . |
| 16 | $XOR_6$ | . | . |
| 17 | $XOR_7$ | . | . |
| . | . | . | . |
| 30 | **$XOR_0 \oplus \ldots \oplus XOR_7$/$Register_{14}$** | . | . |
| 31 | **$XOR_0 \oplus \ldots \oplus XOR_8$/$Register_{15}$** | . | . |

Figure 21. Contents of Frame Buffer *A* after the algorithm terminates after one computation step the new registers values are shown at the specified locations.

### 5.4. Performance Evaluation and Analysis

The algorithm in Table 12 (*CRC-CCITT-16* Parallel Algorithm for a single channel) takes 30 cycles to complete. The speed in bits per cycle of the algorithm of Table 12 is equal to 0.267 bits/cycles i.e. 3.75 cycles for each bit. The time for the algorithm to terminate is equal to 0.3 μsec, and the data rate is 26.67 Mbps.

The algorithm in Table 13 (*CRC-16* Parallel Algorithm for a single channel) takes 26 cycles in order to terminate. The cycle time for the *MorphoSys* is equal to 10 nsec. Thus, the speed in bits per cycle of the algorithm of Table 13 is equal to 0.307 bits/cycles i.e. 3.25 cycles for each bit. The time for the algorithm to terminate is equal to 0.26 μsec, and then the rate in Mega bits per second (Mbps) is 30.76 Mbps.

The *MorphoSys* can calculate in parallel the input of up to 8-channels simultaneously. The results are shown in Table 14.

**Table 14. Table of results.**

| Algorithms | No. of Channels | No. of Cycles | Time in Micro | Bits per Cycle | Mega Bits Per Second | Cycles per Bits |
|---|---|---|---|---|---|---|
| **CRC-CCITT-16** | 1 | 30 | 0.3 | 0.267 | 26.67 | 3.75 |
| **CRC-16** | 1 | 26 | 0.26 | 0.307 | 30.76 | 3.25 |
| **CRC-CCITT-16** | 8 | 30 | 0.3 | 2.13 | 213.13 | 0.46 |
| **CRC-16** | 8 | 26 | 0.26 | 2.46 | 246.15 | 0.41 |

## 6. Conclusion

This chapter has introduced reconfigurable computers as powerful modern supercomputing architectures. The chapter have also investigated the main reasons behind the current advancement in the development of reconfigurable systems. A technical survey of various reconfigurable systems is included laying common grounds for comparisons. In addition, this chapter have presented case studies implemented under the *MorphoSys*. The selected case studies belong to computer graphics and information coding. Parallel versions of the studied algorithms are developed to match the topologies supported by the *MorphoSys*. Performance evaluation and results analyses are included for implementations with different characteristics.